\documentclass[aps,twocolumn,pra,amsmath,amssymb,superscriptaddress,floats]{revtex4}

\usepackage{overpic} 
\usepackage{multirow}

\usepackage{amsmath}	
\usepackage{gensymb}
\usepackage{hyperref}
\hypersetup{colorlinks=true}
\usepackage{upgreek}
\usepackage{graphics}
\usepackage{hyperref}
\usepackage{epsfig}
\usepackage{color}
\usepackage{bm}
\usepackage{siunitx}
\usepackage{bbm}
\usepackage{float}
\usepackage{ulem} 
\usepackage{dsfont}                 
\usepackage{wasysym}

\usepackage{accents}

\usepackage{graphicx}

\graphicspath{{./}{./plots/}}

\usepackage{CJK}
\usepackage{url}
\usepackage{multirow}

\def\ket#1{|#1\rangle }

\def\braket#1{\langle #1 \rangle}

\def\vec{\mathbf}

\newcommand{\upperRomannumeral}[1]{\uppercase\expandafter{\romannumeral#1}}

\begin{document}

\title{Spontaneous Chiral-Spin Ordering in Spin-Orbit Coupled Honeycomb Magnets}
\author{Qiang Luo}
\thanks{Present address: College of Science, Nanjing University of Aeronautics and Astronautics, Nanjing, 211106, China}
\affiliation{Department of Physics, University of Toronto, Toronto, Ontario M5S 1A7, Canada}

\author{P. Peter Stavropoulos}
\affiliation{Department of Physics, University of Toronto, Toronto, Ontario M5S 1A7, Canada}

\author{Jacob S. Gordon}
\affiliation{Department of Physics, University of Toronto, Toronto, Ontario M5S 1A7, Canada}

\author{Hae-Young Kee}
\email[]{hykee@physics.utoronto.ca}
\affiliation{Department of Physics, University of Toronto, Toronto, Ontario M5S 1A7, Canada}
\affiliation{Canadian Institute for Advanced Research, Toronto, Ontario, M5G 1Z8, Canada}
\date{\today}

\begin{abstract}
  Frustrated magnets with highly degenerate ground states are at the heart of hunting exotic states of matter.
  Recent studies in spin-orbit coupled honeycomb magnets have generated immense interest in bond-dependent interactions,
  appreciating a symmetric off-diagonal $\Gamma$ interaction which exhibits a macroscopic degeneracy in the classical limit.
  Here, we study a generic spin model and discover a novel chiral-spin ordering
  with spontaneously broken time-reversal symmetry near the dominant $\Gamma$ region.
  The chiral-spin phase is demonstrated to possess a staggered chirality relation in different sublattices,
  and it exhibits gapless excitations as revealed by the vanishing energy gap and the finite central charge on cylinders.
  Although there is a vestige of a tiny peak in the corner of the second Brillouin zone,
  the magnetic order is likely to vanish as the system size increases.
  Finally, we also attempt to gain insight into the possible topological signature of the chiral-spin phase
  by calculating the dynamic structure factor and the modular $\mathcal{S}$ matrix.
\end{abstract}


\maketitle

\section{Introduction}
Understanding of the emergent phenomena in exotic honeycomb magnets with strong spin-orbit coupling
has triggered an enduring interest in systems that exhibit a novel type of ordering
due to frustration \cite{WKCKB2014,BanerjeeNatMat2016,RauLeeKee2016,TakagiTJ2019,WenYLYL2019}. 
This in turn leads to a particularly high degeneracy of ground state,
which may be lifted by delicate effects such as quantum fluctuation or entropic difference \cite{BergmanAG2007,ShokefSL2011}.
The magnetically ordered states are often selected from their classical degenerate manifolds,
hence quantum spin liquids~(QSLs) are rare and are confined to several specific cases \cite{Anderson1973,Balents2010}.
The Kitaev QSL is a rare example which owns fractionalized excitations of Majorana fermions and gauge fluxes \cite{Kitaev2006}.
Although it has massive classical degenerate ground states,
a finite region of QSL is demonstrated to exist in the presence of other interactions \cite{RanLeeKeePRL2014,KatukuriNJP2014}.

Along the realization of Kitaev model in honeycomb materials like $\alpha$-RuCl$_3$
 \cite{Jackeli2009,YeCCetal2012,ChoiCKetal2012,PlumbCSetal2014,ChunKKetal2015,WangDYL2017,WinterNcom2018},
another bond-dependent $\Gamma$ interaction was identified \cite{RanLeeKeePRL2014},
attracting numerous interest because of the inherently strong frustration \cite{RousPerkins2017,SahaFZetal2019}.
Its classical limit is known as a classical spin liquid with a macroscopic ground-state degeneracy \cite{RousPerkins2017},
which includes the antiferromagnetic~(AFM) phase, zigzag phase, 120$^{\circ}$ phase,
and also large-unit-cell~(LUC) magnetically ordered states \cite{RanLeeKeePRL2014,RousPerkins2017}.
However, quantum ground state of the AFM $\Gamma$ model is more enigmatic,
and a few controversial proposals have been reported,
including a zigzag phase \cite{WangNmdLiu2019}, a nematic paramagnet \cite{GohlkeCKK2020},
and also a disordered state dubbed $\Gamma$ spin liquid ($\Gamma$SL) \cite{CatunYWetal2018,GohlkeWYetal2018,LuoNPJ2021}.

On the other hand, the chiral-spin~(CS) state with spontaneously time-reversal symmetry~(TRS) breaking
was initially proposed over 30 years ago \cite{KalLau1987,WenWlkZee1989},
and has been studied in some geometrically frustrated magnets
\cite{YaoKivelson2007,NielsenSC2013,BauerCKetal2014,GongZS2014,HeShengChen2014,FerrazREP2019,GongZLetal2019,SzaszMZM2020,ChenCW2020,JngJng2020,WangCS2021}.
Within the framework of parton mean-field theory \cite{WenQFT2004}, which is an approximate method and may overestimate the regime of a certain phase,
possible signatures of the CS phase as revealed by nonzero Chern numbers have been presented \cite{RalkoMerino2020,GoJungMoon2019}.
Beyond that, it has not yet been reported convincingly on a honeycomb lattice except by applying TRS breaking terms
such as a magnetic field \cite{LiuNormand2018,RalkoMerino2020} and a three-spin $\hat{\chi}$-interaction \cite{HickeyCP2017,HuangDST2021}.

In this work we study a generic $JK\Gamma\Gamma'$ model including the Heisenberg~($J$) interaction,
Kitaev ($K$) interaction, and symmetric off-diagonal $\Gamma$ and $\Gamma'$ interactions \cite{RanLeeKeePRL2014,RauKeeArXiv2014},
focusing on the AFM $\Gamma$ region.
The scalar spin chirality  \cite{WenWlkZee1989} is used to search for a CS state.
We start by a brief analysis of the chirality on the $\Gamma$ model at the classical level,
followed by a joint exact diagonalization (ED) and density matrix renormalization group~(DMRG) study of the extended Hamiltonian \cite{White1992,Peschel1999,StoudWhite2012}.
Strikingly, we uncover a CS phase where the magnetization tends to vanish after an extrapolation while $\hat{\chi}$ remains be finite.
In addition, it possesses a twofold ground-state degeneracy as a result of the spontaneous TRS breaking.

\begin{figure}[!ht]
\centering
\includegraphics[width=1.00\columnwidth,clip]{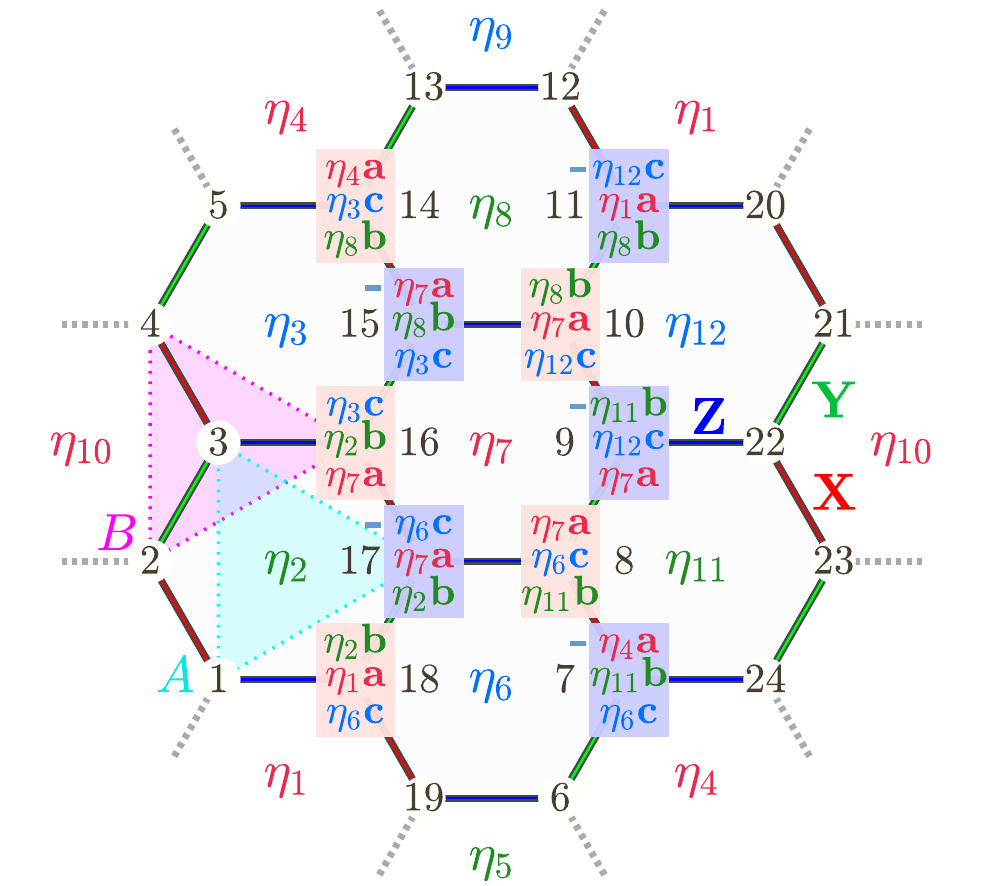}
\caption{Visualization of the classical ground state of the AFM $\Gamma$ model by a set of Ising variables $\{\eta_n\}$ on individual hexagons.
    Here, $\textbf{X}$ (red), $\textbf{Y}$ (green), and $\textbf{Z}$ (blue) represent three different bonds shown in Eq.~\eqref{EQ:JKGGp-Ham},
    and sites of $A$ and $B$ sublattices are labeled by odd and even numbers, respectively.
    The scalar chirality $\chi_n^A$ and $\chi_p^B$ are defined as the shaded cyan and pink triangles, respectively.}
    \label{FIG-HexGam}
\end{figure}

\section{Classical $\Gamma$ model with staggered scalar chirality}
To understand the origin of finite scalar chirality,
we start from a classical $\Gamma$ model which is known to possess a macroscopic ground-state degeneracy.
The spin at site $n$ could be parameterized by
$\hat{\textbf{S}}_n = (\eta_i a,\eta_j b,\eta_k c)$
where ($a$, $b$, $c$) = ($|S^x_n|$, $|S^y_n|$, $|S^z_n|$)
and $\eta_p = \pm 1$ is an Ising variable defined on a hexagon $p$ as shown in Fig.~\ref{FIG-HexGam} \cite{RousPerkins2017}.
The $\eta$ variables can be divided into three types situated in interpenetrating triangular sublattices.
For a $N$-site cluster there are $N/2$ local Ising variables,
leading to a $2^{N/2}$-fold ground-state degeneracy in addition to free choices of $(a, b, c)$
with $\sqrt{a^2 + b^2 +c^2} = S$ \cite{RousPerkins2017}.
The AFM phase and the zigzag phase belong to a special family where $a = b = c = S/\sqrt3$.
Nevertheless, most of the configurations are still the virgin lands with luxuriant noncoplanar phases.

To study the possible CS ordering out of the degenerate manifold, we introduce the scalar chirality \cite{WenWlkZee1989}
\begin{equation}\label{EQ:ChiIJK}
\hat{\chi}^{\triangle}_{ijk} = \hat{\mathbf{S}}_i\cdot(\hat{\mathbf{S}}_j\times\hat{\mathbf{S}}_k),
\end{equation}
where sites ($i, j, k$) form an equilateral triangle~($\triangle$),
which belongs to either $A$ or $B$ sublattice, in the clockwise direction.
Depending on the center of triangle, we distinguish two kinds of $\hat{\chi}$ for each sublattice.
If it is centered at a site (site 3, say) we call it as $\chi_p$,
otherwise it is inside a hexagon and thus termed $\chi_n$, see Fig.~\ref{FIG-HexGam}.
For each hexagon, $\chi_n$ exhibits an intrinsic staggered relation between $A$ and $B$ sublattices
(see Supplemental Material~(SM) \cite{SuppMat} for details).
Considering the hexagon with $\eta_7$ in the center, it is straightforward to check that
the scalar chirality for spins at (9, 17, 15) of the $A$ sublattice is
\begin{align}\label{EQ:ChiAEta7}
\hat{\chi}^A_{9,17,15}
&= \eta_7^3a^3 + \eta_{2}\eta_{8}\eta_{11}b^3 + \eta_{3}\eta_{6}\eta_{12}c^3  \nonumber\\
&\quad\, -abc\eta_7\big(\eta_{3}\eta_{11} + \eta_{2}\eta_{12} + \eta_{6}\eta_{8}\big)
\end{align}
and the scalar chirality for spins at (16, 10, 8) of the $B$ sublattice is
\begin{align}\label{EQ:ChiBEta7}
\hat{\chi}^B_{16,10,8}
&= abc\eta_7\big(\eta_{3}\eta_{11} + \eta_{2}\eta_{12} + \eta_{6}\eta_{8}\big)    \nonumber\\
&\quad\, -\big(\eta_7^3a^3 + \eta_{2}\eta_{8}\eta_{11}b^3 + \eta_{3}\eta_{6}\eta_{12}c^3\big),
\end{align}
which lead to the intrinsic staggered relation
\begin{eqnarray}\label{EQ:ABChi}
\hat{\chi}^A_{9,17,15} = - \hat{\chi}^B_{16,10,8}.
\end{eqnarray}
When CS ordering occurs, i.e.,
$\hat{\chi}^{\triangle}_{ijk} = \hat{\chi}^{\triangle}_{lmn}$ ($\triangle = A, B$),
we find that Ising variables in each triangular sublattices should be equal,
leaving only three flavors of $\eta$s called $\eta_a$ (red), $\eta_b$ (green), and $\eta_c$ (blue).
Under such a restriction, $\chi_p$ also obeys a staggered relation between $A$ and $B$ sublattices.
Furthermore, within each sublattice $\chi_p$ and $\chi_n$ have the \textit{opposite} sign,
namely, $\chi_p^{\triangle} = -\chi_n^{\triangle}$.
While the $(a, b, c)$ degrees of freedom are still left, they are lifted via an order-by-disorder effect,
generating a magnetically ordered state with broken translational symmetry.

To affirm that the magnetic order parameter of the classical CS ordering is finite,
it is naturally to introduce the static structure factor (SSF)
$\mathbb{S}_N({\bf{Q}}) = \sum_{\alpha\beta}\delta_{\alpha\beta}\mathbb{S}_N^{\alpha\beta}({\bf{Q}})$
where
\begin{equation}\label{EQ:SSF}
\mathbb{S}_N^{\alpha\beta}({\bf{Q}})=\frac{1}{N}\sum_{ij}
\langle{S^{\alpha}_i {S^{\beta}_j}}\rangle e^{i{\bf{Q}}\cdot{({\bm{R}}_i-{\bm{R}}_j)}}.
\end{equation}
The order parameter is defined as $M_N({\bf{Q}}) = \sqrt{\mathbb{S}_N({\bf{Q}})/N}$
with ${\bf{Q}}$ being the ordering wavevector.
In the classical CS phase, its SSF could only peak at $\textbf{K}$ point (corner of the first Brillouin zone)
and $\boldsymbol{\Gamma}'$ point (corner of the second Brillouin zone) in the reciprocal space,
as a consequence of the three-sublattice restriction of $\eta$ variables.
In the thermodynamic limit, the magnetic order parameter is defined as $M({\bf{Q}}) = \lim_{N\to\infty} M_N({\bf{Q}})$
with ${\bf{Q}} = \textbf{K}$ or ${\bf{Q}} = \boldsymbol{\Gamma}'$.
After a straightforward calculation it is found that
\begin{align}
M^2(\boldsymbol{\Gamma'}) &= \frac14\big[(a^2\!+\!b^2\!+\!c^2) \!+\! 2(\eta_a\eta_bab \!+\! \eta_b\eta_cbc \!+\! \eta_c\eta_aca)\big] \nonumber\\
&= \frac14\big[(\bar{a}^2+\bar{b}^2+\bar{c}^2) + 2(\bar{a}\bar{b} + \bar{b}\bar{c} + \bar{c}\bar{a})\big]
\end{align}
and
\begin{align}
M^2(\textbf{K}) &= \frac16\big[(a^2\!+\!b^2\!+\!c^2) \!-\! (\eta_a\eta_bab \!+\! \eta_b\eta_cbc \!+\! \eta_c\eta_aca)\big] \nonumber\\
&= \frac16\big[(\bar{a}^2+\bar{b}^2+\bar{c}^2) - (\bar{a}\bar{b} + \bar{b}\bar{c} + \bar{c}\bar{a})\big].
\end{align}
Here, $\bar{a} = \eta_aa$, $\bar{b} = \eta_bb$, and $\bar{c} = \eta_cc$.
Mathematically, for any $\bar{a}, \bar{b}, \bar{c} \in[-S, S]$ and $\bar{a}^2+\bar{b}^2+\bar{c}^2 = S^2$, one can prove that
$-S^2/2 \leq \bar{a}\bar{b} + \bar{b}\bar{c} + \bar{c}\bar{a} \leq S^2$.
It is worth to note that equality on the left-hand side occurs when $\bar{a}+\bar{b}+\bar{c}=0$,
while equality on the right-hand side holds if and only if $\bar{a}=\bar{b}=\bar{c}=\pm S/\sqrt3$.
Taken together, it is easy to check that
$M(\boldsymbol{\Gamma'}) \in [0, \sqrt3S/2]$ and $M(\textbf{K}) \in [0, S/2]$,
and they can not be zero \textit{simultaneously}.
Hence, the finite magnetization of the classical CS phase implies that it is a magnetically ordered state.

As we delve into the phase region away from the pure $\Gamma$ limit,
we encounter two remarkable sources of quantum fluctuations
stemming from huge ground-state degeneracy and competing interactions.
While the inherent staggered relations for $\chi_p$ and $\chi_n$ within each sublattice should still hold,
other constraints could be eased because of strong quantum fluctuations,
evoking another quantum CS ordered phase without magnetic ordering.
With this in mind, we proceed with ED+DMRG computations to look for numerical evidences.

\section{Model Hamiltonian and Phase Diagrams}
\subsection{Model and methods}
We demonstrate our idea with a paradigmatic spin-$1/2$ $JK\Gamma\Gamma'$ model whose Hamiltonian reads \cite{RanLeeKeePRL2014,RauKeeArXiv2014}
\begin{align}\label{EQ:JKGGp-Ham}
\mathcal{H} =
    & \sum_{\left<ij\right>\parallel\gamma} \Big[J \hat{\mathbf{S}}_i \cdot \hat{\mathbf{S}}_j + K S_i^{\gamma} S_j^{\gamma}
    + \Gamma \big(S_i^{\alpha}S_j^{\beta}+S_i^{\beta}S_j^{\alpha}\big)\Big]   \nonumber \\
    & + \Gamma' \sum_{\left<ij\right>\parallel\gamma}
        \Big[\big(S_i^{\alpha} + S_i^{\beta}\big) S_j^{\gamma} + S_i^{\gamma} \big(S_j^{\alpha} + S_j^{\beta}\big) \Big],
\end{align}
where $S_i^{\gamma}$~($\gamma = x,~y,~z$)
is the $\gamma$-component of spin-1/2 operator at site $i$.
On $\textbf{Z}$ bonds $(\alpha, \beta, \gamma) = (x, y, z)$,
with cyclic permutation for $\textbf{X}$ and $\textbf{Y}$ bonds.
The model~\eqref{EQ:JKGGp-Ham} is studied in the vicinity of the $\Gamma$ limit where $\Gamma = 1$ is set as the energy unit.
For simplicity, we consider diagonal interactions $J$ and $K$ on equal footing, i.e., $J = K$.
The classical phase diagram is obtained by the Luttinger-Tisza analysis \cite{ChalKhal2015}
and classical Monte Carlo method \cite{Metropolis1953,HukushimaNemoto1996}
(see Sec.~I in the SM \cite{SuppMat} for details).
The linear spin-wave theory (LSWT) analysis is also utilized when necessary.
Depending on the circumstances, three kinds of geometries are employed in DMRG calculations \cite{White1992,Peschel1999,StoudWhite2012}.
Firstly, we focus primarily on a $C_3$-symmetric hexagonal cluster of $N$ = 24 with periodic boundary conditions
illustrated in Fig.~\ref{FIG-HexGam}.
During the calculation, we keep as many as $m$ = 3000 block states
and execute up to 12 sweeps until the largest truncation error is smaller than $10^{-6}$.
Secondly, we also adopt the YC$n$ cylinder of $L_x\times L_y$, which is open along the $L_x$ direction and periodic along the $L_y$ direction.
The letter ``Y" means that one of the bonds is along the perpendicular ($L_y$) direction,
while the letter $n$ represents the number of sites along this direction.
Finally, we also consider hexagonal cylinders which are open and periodic
in the $\mathbf{e}_1 (\sqrt3, 0)$ and $\mathbf{e}_2 (\sqrt3/2, 3/2)$ directions, respectively.
As each unit cell contains two sites, the total number of sites is $N = 2L_xL_y$
where $L_x$~($L_y$) represents the number of unit cells along the $\textbf{e}_1$~($\textbf{e}_2$) direction
(see inset of Fig.~\ref{FIG-ChiVNE}(b)).

\begin{figure*}[htb]
\centering
\includegraphics[width=0.95\linewidth, clip]{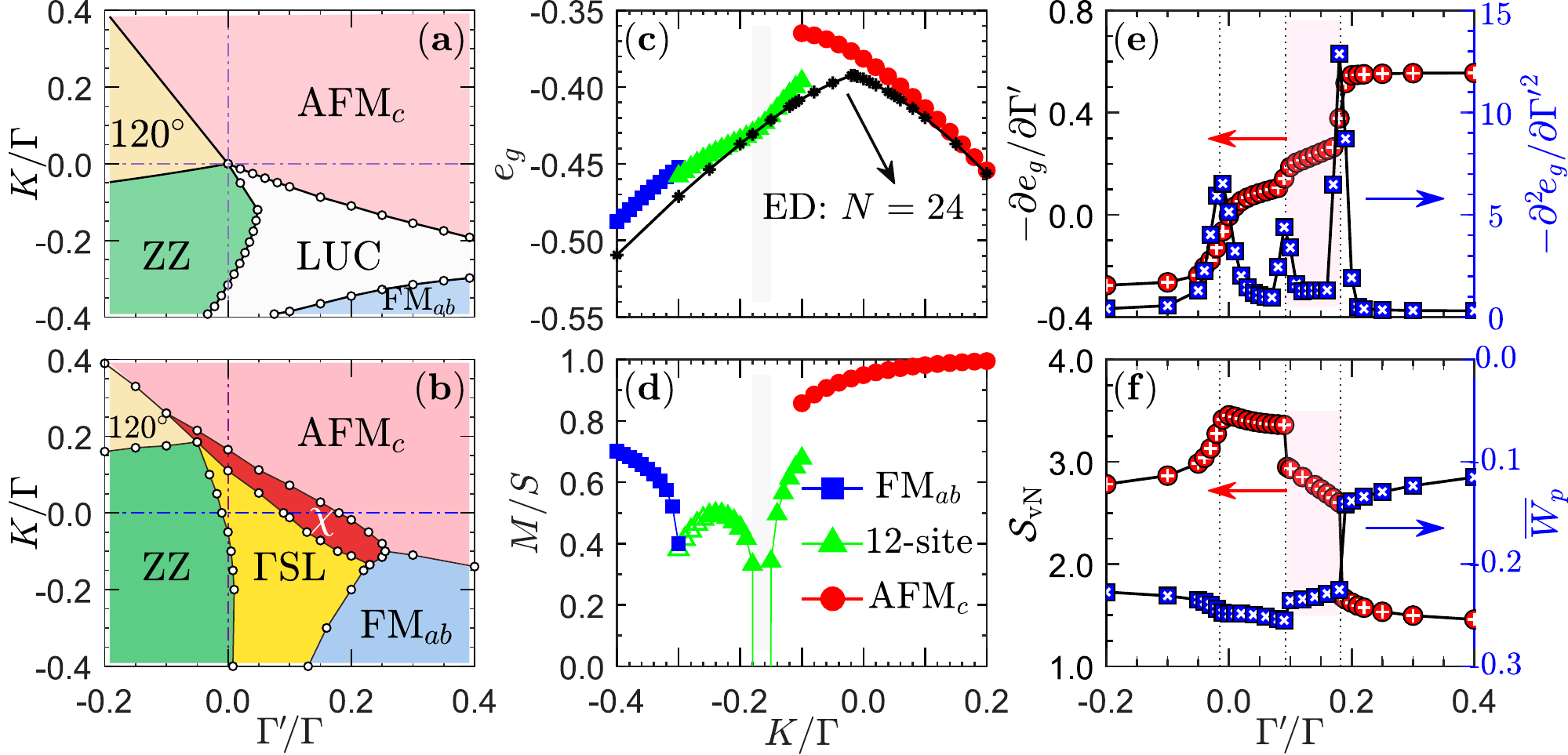}\\
\caption{The (a)~classical and (b)~quantum phase diagrams of model~\eqref{EQ:JKGGp-Ham} in the neighbor of the $\Gamma$ limit.
    The white region in panel (a) contains many LUC phases and also incommensurate phases near phase boundaries.
    Details are shown in the SM \cite{SuppMat}.
    In panel (b), there is a wide region of $\Gamma$SL and a CS phase marked by $\chi$.
    Panels (c)-(d) show the spin-wave energy $e_g$ and magnetization $M$ of the FM$_{ab}$ phase (blue square), 12-site phase (green triangle),
    and AFM$_c$ phase (red circle), along the line of $\Gamma'/\Gamma = 0.2$.
    In panel (c), the ED energy on a hexagonal cluster of $N$ = 24 is shown for comparison (black asterisk).
    In panel (d), the open symbol (green triangle) represents the extrapolated magnetization where LSWT fails.
    The gray shadow marks the region where the energy $e_g$ is extremely close to the quantum result (black asterisk) and the magnetization $M$ diverges.
    Panels (e)-(f) focus on the $\Gamma$-$\Gamma'$ limit in the window of $-0.2 \leq \Gamma'/\Gamma \leq 0.4$.
    (e) Behaviors of the first-order (red circle) and second-order (blue square) ground-state energy derivatives.
    (f) Behaviors of the von Neumann entropy $\mathcal{S}_{\rm vN}$ (red circle) and flux-like density $\overline{W}_p$ (blue square).
    }\label{FIG-GGpHKPD}
\end{figure*}

\subsection{Classical phase diagram}

We begin by mapping out the classical phase diagram of model~\eqref{EQ:JKGGp-Ham} in the vicinity of the $\Gamma$ limit.
As shown in Fig.~\ref{FIG-GGpHKPD}(a), apart from a FM$_{ab}$ phase (blue) in the right-bottom corner,
three magnetically ordered states, the AFM$_c$ phase (pink), the zigzag phase (green), and the $120^{\circ}$ phase (yellow),
are selected from the degenerate ground state of the classical $\Gamma$ model \cite{RousPerkins2017}.
The subscripts imply that the classical moment direction of the FM$_{ab}$ phase lies in the honeycomb plane,
while it is along the out-of-plane direction for the AFM$_c$ phase.
In addition, the phase diagram also contains dozens of LUC phases and a few incommensurate phases (shown in the white region).
The classical energy of the conventional magnetically ordered states (i.e., AFM$_c$, 120$^{\circ}$, zigzag, and FM$_{ab}$)
could be given analytically by the Luttinger-Tisza method.
This method is powerful for the determination of classical magnetic ground states,
and its application to the spin-orbit coupled model has been outlined in a previous work \cite{ChalKhal2015}.
Within the framework of the Luttinger-Tisza analysis, we find that the classical energy per site of the AFM$_c$ phase is
\begin{equation}\label{EclNeel}
\varepsilon_{\mathrm{cl}}^{\textrm{AFM$_c$}} = -\big(\Gamma+2\Gamma'+2K\big),
\end{equation}
and the energy of the spiral $120^{\circ}$ phase is
\begin{equation}\label{Ecl120}
\varepsilon_{\mathrm{cl}}^{120^{\circ}} = -\frac{1}{2}\big[2(\Gamma-\Gamma')+K\big].
\end{equation}
The $120^{\circ}$ phase is not favored by the FM Kitaev interaction,
which then gives way to the zigzag phase whose energy is
\begin{align}\label{EclZigzag}
\varepsilon_{\mathrm{cl}}^{\textrm{zigzag}} = -\frac{1}{4}\big(\mathcal{R} + \sqrt{8\Gamma^2+\mathcal{R}^2}\big)
\end{align}
where $\mathcal{R} = \Gamma-2\Gamma'-2K$.
The classical moment direction of the zigzag phase is tilted away from the $ab$ plane in the crystalline reference frame
and the tilted angle is known to depend on the interaction parameters \cite{ChalKhal2015}.

As can be seen from Fig.~\ref{FIG-GGpHKPD}(a), in the left part of the phase diagram where $\Gamma' < 0$,
the $120^{\circ}$ phase is sandwiched between the AFM$_c$ phase and the zigzag phase.
For the AFM$_c$--$120^{\circ}$ transition, the transition boundary is
\begin{equation}\label{EQ:TranAFvs120}
K_{c,u} = -2\Gamma',
\end{equation}
while for the zigzag--$120^{\circ}$ transition it is
\begin{equation}\label{EQ:TranZZvs120}
K_{c,l} = \frac{\sqrt{49\Gamma^2-60\Gamma\Gamma'+36\Gamma'^2}-(7\Gamma-6\Gamma')}{6}.
\end{equation}
Specifically, when $\vert\Gamma'/\Gamma\vert \ll 1$ Eq.~\eqref{EQ:TranZZvs120} is reduced to
\begin{equation}\label{EQ:TranZZvs120Asy}
K_{c,l} = \frac27\Big[\Gamma' + \Big(\frac{6}{7}\Gamma'\Big)^2\Big] + \mathcal{O}\big(\Gamma'^3\big).
\end{equation}
On the other side where $\Gamma' > 0$, the AFM$_c$ phase occupies the majority region of the phase diagram.
However, for large enough FM Kitaev interaction, there is a FM$_{ab}$ phase whose energy is
\begin{equation}\label{EclFMab}
\varepsilon_{\mathrm{cl}}^{\textrm{FM}_{ab}} = -\frac{1}{2}(\Gamma+2\Gamma') + 2K.
\end{equation}
One should notice that there is not a direct transition between the AFM$_c$ phase and the FM$_{ab}$ phase for modest $\Gamma'$.
Instead, many magnetically ordered LUC states are demonstrated to exist in between.
These LUC phases are unstable against quantum fluctuation,
providing a fertile playground for the realization of disordered phases at the quantum level \cite{LeeKCetal2020}.

\subsection{Spin-wave analysis in the LUC region}\label{SUBSEC:LSWT}
The simplest way to roughly demonstrate that LUC orderings could be melted by quantum fluctuations is perhaps the LSWT,
which is an efficient method that has been applied successfully in various spin-orbit coupled models on the honeycomb lattice
\cite{ChoiCKetal2012,WinterNcom2018,ChernKLK2020}.
In this method, the spins are rewritten as the bosonic creation and annihilation operators,
and the Hamiltonian $\mathcal{H}$ is reduced to a quadratic form $\mathcal{H}_{\rm sw}$ up to $\mathcal{O}(1/S)$.
The spin-wave dispersion $\omega_{\bm{q}\upsilon}$ and the corresponding eigenvector $\vec{v}_{\bm{q}}^{(\upsilon)}$ are readily obtained.
Here, $\bm{q}$ is the wavevector in the reciprocal space and $\upsilon = 1,2,\cdots,n_s$, with $n_s$ being the number of sites in the unit cell.
As a result, the spin-wave energy $e_g$ is given by
\begin{eqnarray}
e_g = S(S+1) \varepsilon_{\textrm{cl}} + \frac{S}{2n_s} \sum_{\{\upsilon\}\in n_s} \int \frac{d^2\bm{q}}{(2\pi)^2} \omega_{\bm{q}\upsilon},
\end{eqnarray}
and the classical moment $M$ is
\begin{eqnarray}
\frac{M}{S} = 1 - \frac{1}{n_sS} \sum_{\{\upsilon\}\in n_s} \int \frac{d^2\bm{q}}{(2\pi)^2} \left|\vec v_{-\bm{q}}^{*(\upsilon)}\right|^2.
\end{eqnarray}

To get started, it is useful to recall the classical Monte Carlo result of the model in Eq.~\eqref{EQ:JKGGp-Ham}.
For this purpose, let us focus on the transitions along the line of $\Gamma' = 0.20$,
which is known to host a FM$_{ab}$ phase when $K < -0.34$ and an AFM$_c$ phase when $K > -0.11$
(for details, see Fig. 2 in the SM \cite{SuppMat}).
In between, there are three LUC phases whose magnetic unit cells are 44-site, 12-site, and 24-site, respectively, with the increase of $K$.
Within the LUC region, the extent of the remaining two is considerably smaller than the middle 12-site phase and thus are ignored tentatively.
Further, the classical magnetic moment directions are essential for the spin-wave analysis.
For the AFM$_c$ phase it is simply along the \textbf{c}[111] direction.
By contrast, for the FM$_{ab}$ phase the spins are free to rotate in the $ab$ plane as a consequence of an emergent $U(1)$ symmetry.
However, this continuous $U(1)$ symmetry is fragile against the quantum fluctuation,
and breaks to a discrete $\mathbb{Z}_2$ symmetry with a energetically favored moment direction along the \textbf{b}[$\bar1$10] axis.
We stress that such an order-by-disorder effect has also been explored in the stripe phase \cite{LuoNPJ2021,ZhuMakWhiteetal2017}.
The magnetic moment of the 12-site phase is more complicated,
but it can be determined by the Monte Carlo method or the energy optimization method illustrated in the SM \cite{SuppMat}.

The LSWT calculations of the energy $e_g$ and magnetization $M$ on the FM$_{ab}$ phase (blue square), 12-site phase (green triangle),
and AFM$_c$ phase (red circle) are shown in Fig.~\ref{FIG-GGpHKPD}(c) and Fig.~\ref{FIG-GGpHKPD}(d), respectively.
In Fig.~\ref{FIG-GGpHKPD}(c), the ED energy on a hexagonal cluster of $N$ = 24 is shown for comparison.
In the whole parameter region, the spin-wave energy $e_g$ is higher than the quantum result (black asterisk),
and there is a magnetization reduction in the magnetically ordered phases, which is more pronounced in the 12-site phase.
However, the most striking observation is that in the narrow region of $-0.18 \leq K \leq -0.15$,
the spin-wave energy (green triangle) is extremely close to the ED energy (black asterisk).
In addition, we find that the magnetization $M$ diverges as it goes to a (infinite) negative value,
indicating a strong quantum fluctuation which melts the magnetic ordering.
Notably, this region is consistent with the $\Gamma$SL phase in the quantum phase diagram, cf. Fig.~\ref{FIG-GGpHKPD}(b).
Thus, our LSWT results suggest the the LUC phases are unstable against quantum fluctuations, giving rise to the disordered phases in certain region.

\begin{figure}[!ht]
\centering
\includegraphics[width=1.00\columnwidth, clip]{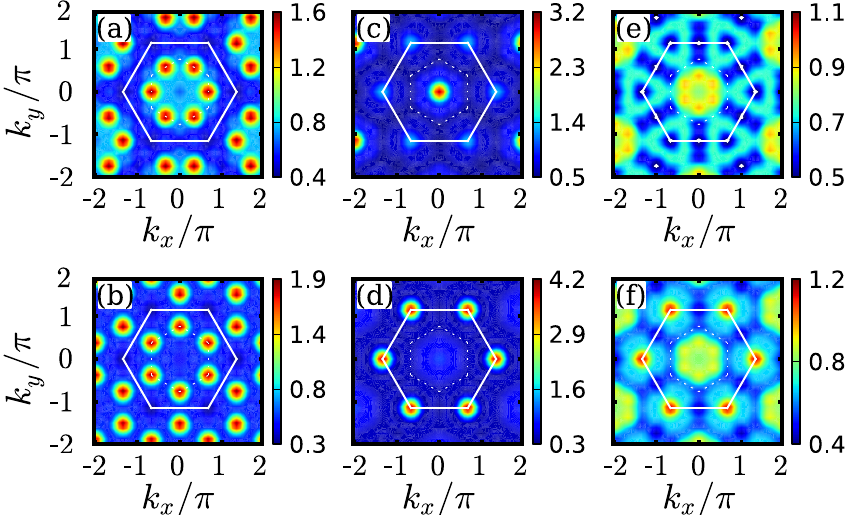}\\
\caption{Typical contour plots of the SSFs in the phase diagram shown in Fig.~\ref{FIG-GGpHKPD}(b), which includes
  (a) the zigzag phase        with $(K, \Gamma, \Gamma')$ = $(0.0, 1.0, -0.1)$,
  (b) the 120$^{\circ}$ phase with $(K, \Gamma, \Gamma')$ = $(0.3, 1.0, -0.2)$,
  (c) the FM$_{ab}$ phase     with $(K, \Gamma, \Gamma')$ = $(-0.2, 1.0, 0.3)$,
  (d) the AFM$_c$ phase       with $(K, \Gamma, \Gamma')$ = $(0.0, 1.0, 0.3)$,
  (e) the $\Gamma$SL          with $(K, \Gamma, \Gamma')$ = $(0.0, 1.0, 0.05)$,
  and (f) the CS phase        with $(K, \Gamma, \Gamma')$ = $(0.0, 1.0, 0.15)$.
  }\label{FIG-SSF}
\end{figure}

\subsection{Quantum phase transitions}

\begin{figure}[!ht]
\centering
\includegraphics[width=0.95\columnwidth, clip]{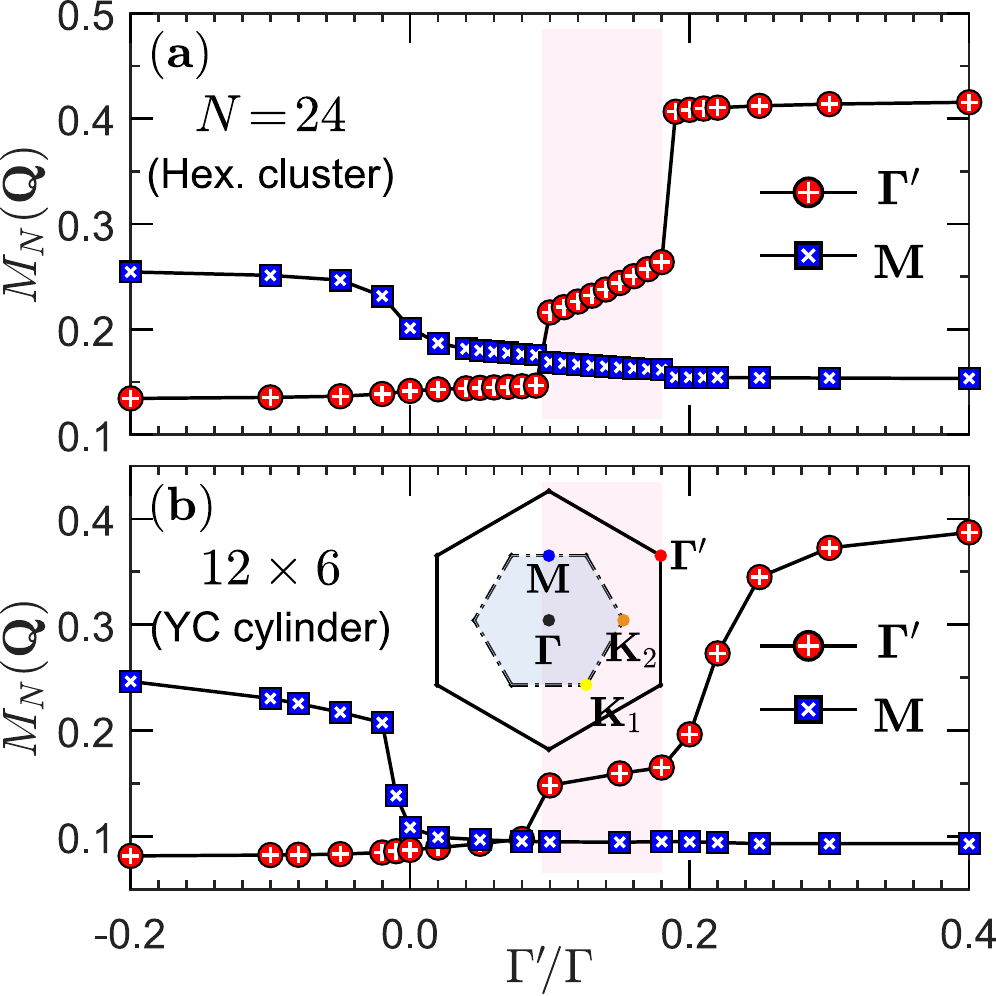}\\
\caption{Order parameters $M_N(\textbf{Q})$ for the zigzag phase (red square) and AFM$_c$ phase (blue circle)
    with $\textbf{Q} = \textbf{M}$ and $\boldsymbol{\Gamma}'$, respectively, in the $\Gamma$-$\Gamma'$ limit.
    Panels (a) and (b) are results obtained on the 24-site hexagonal cluster and YC6 cylinder of $12\times6$, respectively.
    The inset of panel (b) shows the Brillouin zone with high-symmetry points.}
    \label{FIG-GGpOP}
\end{figure}

The quantum phase diagram is shown in Fig.~\ref{FIG-GGpHKPD}(b),
which contains six distinct phases and four of them are magnetically ordered as identified in the classical case.
The typical SSFs $\mathbb{S}(\textbf{Q})$ of the four magnetically ordered phases on a 24-site hexagonal cluster
are illustrated in Fig.~\ref{FIG-SSF}(a,b,c,d).
It is clearly shown that the zigzag phase (panel a), the 120$^{\circ}$ phase (panel b), the FM$_{ab}$ phase (panel c), and the AFM$_c$ phase (panel d)
display the ordering wavevectors at $\textbf{M}$ point, $\textbf{K}$ point, $\boldsymbol{\Gamma}$ point, and $\boldsymbol{\Gamma}'$ point in the Brillouin zone,
respectively (for the illustration of the high-symmetry points, see inset of Fig.~\ref{FIG-GGpOP}(b)).

Throughout the phase diagram, it also acquires two emergent phases which do not have a classical analogy.
One is the $\Gamma$SL that can be compatible with a small positive $\Gamma'$ interaction.
It is named after the ground state of pure $\Gamma$ model that is proposed to be a gapless QSL  \cite{LuoNPJ2021}.
As shown in Fig.~\ref{FIG-SSF}(e), there are only soft peaks in the reciprocal space,
and these peaks are expected to be more diffusive with the increase of the system size.
The other is a CS ordered phase that appears in a narrow slot between the $\Gamma$SL and the AFM$_c$ phase.
The $\mathbb{S}(\textbf{Q})$ is diffusive inside the first Brillouin zone,
while a tiny peak at the $\boldsymbol{\Gamma}'$ point appears in the second Brillouin zone [see Fig.~\ref{FIG-SSF}(f)].
We note that such a peak is an indication of the TRS breaking.

Below we enumerate the methods to determine the phase boundaries and identify the nature of phases therein.
Without loss of generality, we consider the $\Gamma$-$\Gamma'$ line to focus on the $\Gamma$SL and CS phase.
The basic quantity is the ground-state energy $e_g = E_g/N$ and its derivatives, which exhibit jump or peak at the transition point,
depending on the order of the phase transition.
As shown in Fig.~\ref{FIG-GGpHKPD}(e), the phase loci are pinpointed by the first- and second-order derivatives of the ground-state energy.
There are three consecutive phase transitions at $\Gamma' \approx -0.015$, 0.095, and 0.185, respectively.
Further, we can use the von Neumann entanglement entropy $\mathcal{S}_{\rm vN}$ to distinguish different phases.
It is defined as $\mathcal{S}_{\rm vN} = -\mathrm{tr}(\rho_s\ln\rho_s)$
where $\rho_s$ is the reduced density matrix obtained by tracing over the degrees of freedom in the other half subsystem.
Generally, the transition could be revealed by its turning point or pinnacle/jump position.
Moreover, we can define a hexagonal plaquette operator
$\hat{W}_p = 2^6 S_1^{x}S_2^{y}S_3^{z}S_4^{x}S_5^{y}S_6^{z}$ on a plaquette $p$,
and then introduce the flux-like density $\overline{W}_p = \sum_p\langle\hat{W}_p\rangle/N_p$ where
$N_p = N/2$ is the number of hexagonal plaquettes.
Over the years, the $\overline{W}_p$ has emerged as a sensitive probe to capture quantum phase transitions in a variety of systems
\cite{KatukuriNJP2014,GordonCSetal2019,LuoNPJ2021,LuoBAKG2021}.
Figure~\ref{FIG-GGpHKPD}(f) shows the behaviors of the entanglement entropy $\mathcal{S}_{\rm vN}$ (red circle) and flux-like density $\overline{W}_p$ (blue square),
and it is found that they yield the same transition points to that of the energy derivatives.
Remarkably, the sharp jumps in $\mathcal{S}_{\rm vN}$ and $\overline{W}_p$ at the left and right boundaries of the CS phase
indicate that the transitions therein are of first order.

After having determined the transition points, here we go into the phases via the SSFs.
For $\Gamma' \lesssim -0.015$, the SSF $\mathbb{S}_N({\bf{Q}})$ displays distinct peaks at $\textbf{M}$ point in the first Brillouin zone,
while for $\Gamma' \gtrsim 0.185$ dominating peaks located at $\boldsymbol{\Gamma}'$ point in the second Brillouin zone appear.
These results advocate the semi-classical analysis that AFM $\Gamma'$ interaction favors the AFM$_c$ phase
while even a tiny FM $\Gamma'$ interaction could stabilize the zigzag phase \cite{ChernKLK2020,GordonCSetal2019}.
The order parameters $M_N(\textbf{Q})$ for the zigzag phase (blue square) and the AFM$_c$ phase (red circle) are shown in Fig.~\ref{FIG-GGpOP}(a),
and two intermediate phases are found in between.
One is recognized as the $\Gamma$SL since the SSF is diffusive with soft peaks
and it incorporates the ground state of the $\Gamma$ model which is proposed to be a gapless QSL \cite{LuoNPJ2021}.
The other is referred to as the CS ordered phase whose nature will be clarified later.

\begin{figure}[!ht]
\centering
\includegraphics[width=0.95\columnwidth, clip]{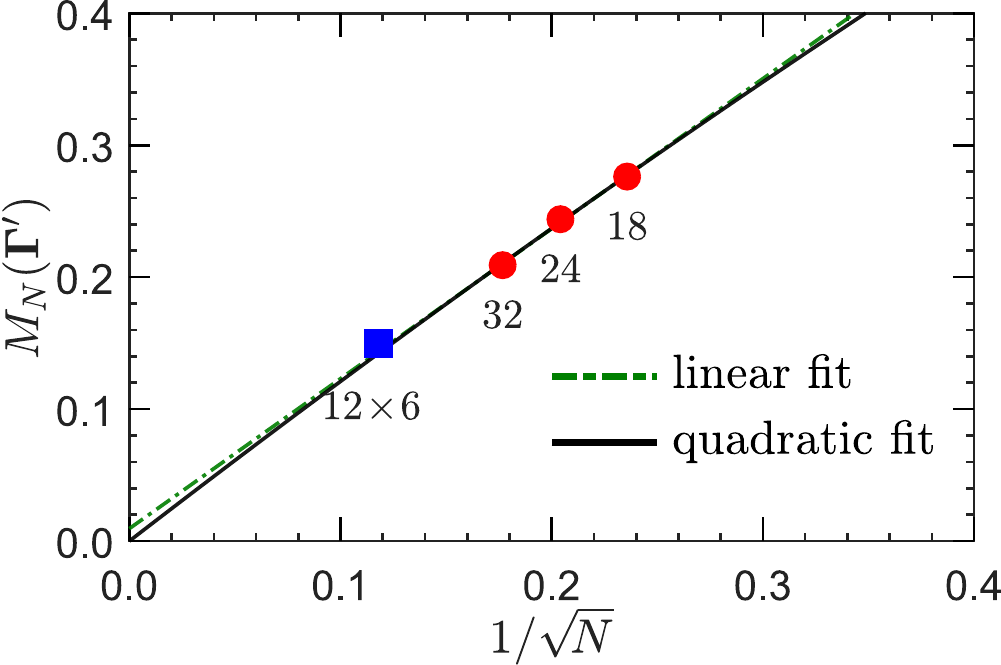}\\
\caption{Extrapolation of $M_N(\boldsymbol{\Gamma'})$ at $\Gamma'/\Gamma = 0.15$ under hexagonal clusters of $N$ = 18, 24, and 32 (red circle).
  The blue dash-dot line represents the linear fitting while the black solid line stands for the quadratic fitting.
  Magnetic order under a $12\times6$ cylinder with $N = 72$ (blue square) is plotted for comparison.
  }\label{FIG-ChiOP}
\end{figure}

To underpin the conclusions drawn from the calculations on the 24-site cluster under full periodic boundary conditions,
we next present the results on the YC6 cylinder,
which strikes a balance between bidimensionality and capacity to converge the DMRG calculation.
We note that the $12\times6$ YC cylinder represents a superior geometry
since the number of sites contained is the multiple of 2, 4, 6, 12, and 18,
which could accommodate with many conventional and LUC orderings and thus reduces the artificial effect.
The order parameters of the zigzag phase and the AFM$_c$ phase on the YC6 cylinder of $12\times6$ are shown in Fig.~\ref{FIG-GGpOP}(b),
from which we can tell that the transition points are almost the same to those of the 24-site case,
corroborating the robustness of the two emergent phases.

In the CS ordered phase, the soft peak at $\boldsymbol{\Gamma}'$ point in the reciprocal space is reminiscent of the TRS breaking.
However, magnitude of the peak decreases with the growing of system size.
Figure~\ref{FIG-ChiOP} shows the extrapolation of the magnetic order parameter $M_N(\boldsymbol{\Gamma'})$ at $\Gamma' = 0.15$ on hexagonal clusters of $N$ = 18, 24, and 32,
and the value of the magnetic order parameter on a $12\times6$ cylinder is also plotted for comparison.
It is found that $M_N(\boldsymbol{\Gamma'})$ shows a roughly linear decrease with $1/\sqrt{N}$,
with an estimated value of 0.01 (0.00) for the linear (quadratic) fitting when $N\to\infty$,
signifying the absence of long-range magnetic ordering in the ground state.
These results suggest that the CS phase is likely a disordered phase,
which stands out in remarkable contrast to the classical CS ordering which are characterized by a nonzero magnetization.

\section{The chiral-spin ordered state}
\subsection{Degeneracy and staggered chirality}
\begin{figure}[!ht]
\centering
\includegraphics[width=0.95\columnwidth, clip]{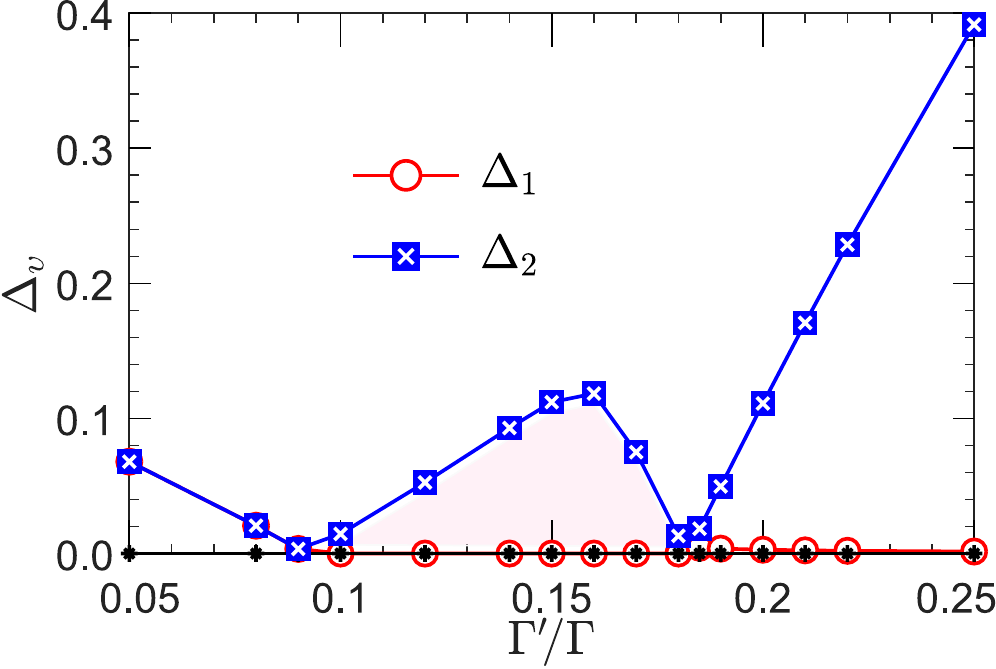}\\
\caption{The first (red circle) and second (blue square) energy gaps $\Delta_{1,2}$ in the range of $0.05 \leq \Gamma'/\Gamma \leq 0.25$.
    The pink region marks the CS phase, which has a finite energy gap in the 24-site cluster.
    }\label{FIG-ChiGap}
\end{figure}

In this section we adopt the momentum-resolved ED calculation and the large-scale DMRG calculation, in conjunction with the symmetry analysis,
to decipher the nature of the CS ordered phase.
Without loss of generality, we proceed to focus on the $\Gamma$-$\Gamma'$ line where the CS phase is sandwiched between the $\Gamma$SL and the AFM$_c$ phase.
Before studying the excitations of the CS phase,
it is useful to recall that the $\Gamma$SL is gapless albeit with a small energy gap retained in finite-size systems \cite{LuoNPJ2021},
and the AFM$_c$ phase is gapped with a doubly degenerate ground state.
Figure~\ref{FIG-ChiGap} shows the first two energy gaps $\Delta_{1,2} = E_{1,2}-E_g$ on a 24-site cluster in the $\Gamma$-$\Gamma'$ model.
The intervening CS phase has a twofold degenerate ground state,
which is well separated from the excited states with a finite energy gap $\Delta_2$ due to the small system size.
Under the 24-site hexagonal cluster, the doubly degenerate ground state is protected by the combination of TRS $\hat{\mathcal{T}}$, inversion symmetry $\hat{\mathcal{P}}$,
and $\hat{C}_{2b}$ about the crystallographic $\hat{\textbf{b}}$ direction.
To target a purified ground state without breaking TRS abruptly, we add a tiny pinning field $h_z = 10^{-3}$
with opposing sign at endpoint sites (i.e., site 1 and site 24 shown in Fig.~\ref{FIG-HexGam}) which belong to different sublattices.
The energy splitting is approximately of $\mathcal{O}(h_z^2)$,
which is far less than the corresponding energy gap $\Delta_2$ and thus only slightly perturbs the ground state.
After locking a specific one, we gradually reduce the strength of $h_z$ and do not stop sweeping
until $h_z$ is vanishingly small. 

Apart from the 24-site case, we have also checked the ground-state degeneracy of the CS phase on 18-site and 32-site hexagonal clusters.
Table~\ref{TAB-ChiEv} shows the first five energy levels on the three clusters where $\Gamma' = 0.15$ is taken as an example.
Apparently, the ground-state degeneracy is always twofold in these cases.
However, the energy gap $\Delta_2$ is not monotonously varying with the increase of the system size.
The energy gap is 0.012, 0.112, and 0.0002 for hexagonal clusters of $N$ = 18, 24, and 32, respectively,
suggesting a likely vanishing energy gap for large enough systems.
The existence of the gapless excitations is also borne out by the entanglement entropy scaling that will be shown later.

\begin{table}[th!]
\caption{\label{TAB-ChiEv}
  The first five energy levels $E_{\upsilon}$ ($\upsilon$ = 0-4) at $\Gamma'/\Gamma$ = 0.15 in the $\Gamma$-$\Gamma'$ model on hexagonal clusters of $N$ = 18, 24, and 32.
  The energy gap is defined as $\Delta_2 = E_{2}-E_{0}$.}
  \begin{ruledtabular}
  \begin{tabular}{ c  c  c  c}
  $\;$        & $N = 18$          & $N = 24$          & $N = 32$  \\
  \colrule
  $E_{0}$     & -6.6497245386     & -9.0381902795     & -12.06667 \\
  $E_{1}$     & -6.6497245386     & -9.0381902795     & -12.06667 \\
  \colrule
  $E_{2}$     & -6.6376518389     & -8.9261211865     & -12.06655 \\
  $E_{3}$     & -6.6099455586     & -8.8509767453     & --        \\
  $E_{4}$     & -6.5769370525     & -8.8509767453     & --        \\
  \colrule
  $\Delta_2$  &  0.0120726997     & 0.1120690930      & 0.00012   \\
  \end{tabular}
  \end{ruledtabular}
\end{table}

\begin{figure}[!ht]
\centering
\includegraphics[width=0.95\columnwidth, clip]{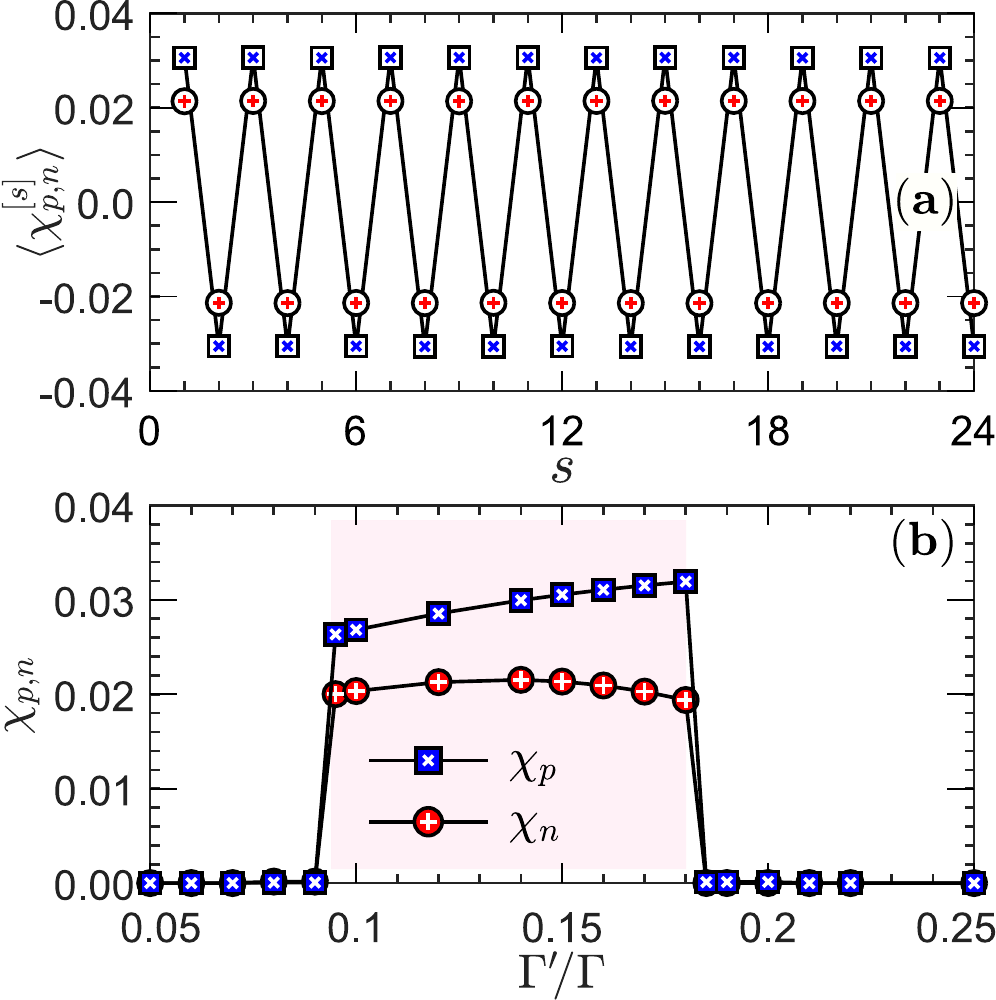}\\
\caption{(a) Chiral distribution of $\chi_p^{[s]}$ (blue square) and $\chi_n^{[s']}$ (red circle) at $\Gamma'/\Gamma$ = 0.15 in the $\Gamma$-$\Gamma'$ model.
    The site number $s$ is shown in Fig.~\ref{FIG-HexGam}, while $s'$ denotes $\eta$'s with a same $\chi_n^B$-$\chi_n^A$ pattern as $\chi_p$.
    (b) The two kinds of chirality $\chi_p$ (blue square) and $\chi_n$ (red circle) in the range of $0.05 \leq \Gamma'/\Gamma \leq 0.25$.
    }\label{FIG-Chirality}
\end{figure}

Next, we turn to unveil the unusual restrictions on the scalar spin chirality in the CS phase.
For any purified ground state, the spatial distributions of $\chi_p$ (blue square) and $\chi_n$ (red circle)
on a 24-site cluster at $\Gamma' = 0.15$ are shown in Fig.~\ref{FIG-Chirality}(a).
For either type of the chirality, there are sign differences between $A$ and $B$ sublattices, in accordance with Eq.~\eqref{EQ:ABChi}.
Intriguingly, the signs of $\chi_p$ and $\chi_n$ are identical within each sublattices and their magnitudes are unequal.
We emphasize that these properties are distinct from the the classical CS ordering stemming from the degenerate ground state of the classical $\Gamma$ model,
indicating the vital role played by $\Gamma'$ interaction.
Pertinently, effects of a small AFM $\Gamma'$ term in honeycomb magnets have been studied recently \cite{MaksimovCherny2020,AndradeJV2020}.
On the other hand, the other degenerate ground state has an opposing sign structure of the scalar chirality,
and it is zero for excited states \cite{SuppMat}.
Figure~\ref{FIG-Chirality}(b) shows the magnitudes of $\chi_p$ and $\chi_n$ in the region of $\Gamma'\in[0.05, 0.25]$,
highlighting the CS phase characterized by a finite chirality of roughly 0.03 (for $\chi_p$) or 0.02 (for $\chi_n$).
We also calculate the chiral distribution on hexagonal clusters of $N$ = 18 and 32 at $\Gamma' = 0.15$,
and we find that the chirality is fairly stable with a negligible finite-size effect.
To summarize, our study on the magnetic order parameters and the scalar chirality leads us to the conclusion
that the CS phase is a magnetically disordered phase but with a rank-2 chiral ordering.

\begin{figure*}[htb]
\centering
\includegraphics[width=0.95\linewidth, clip]{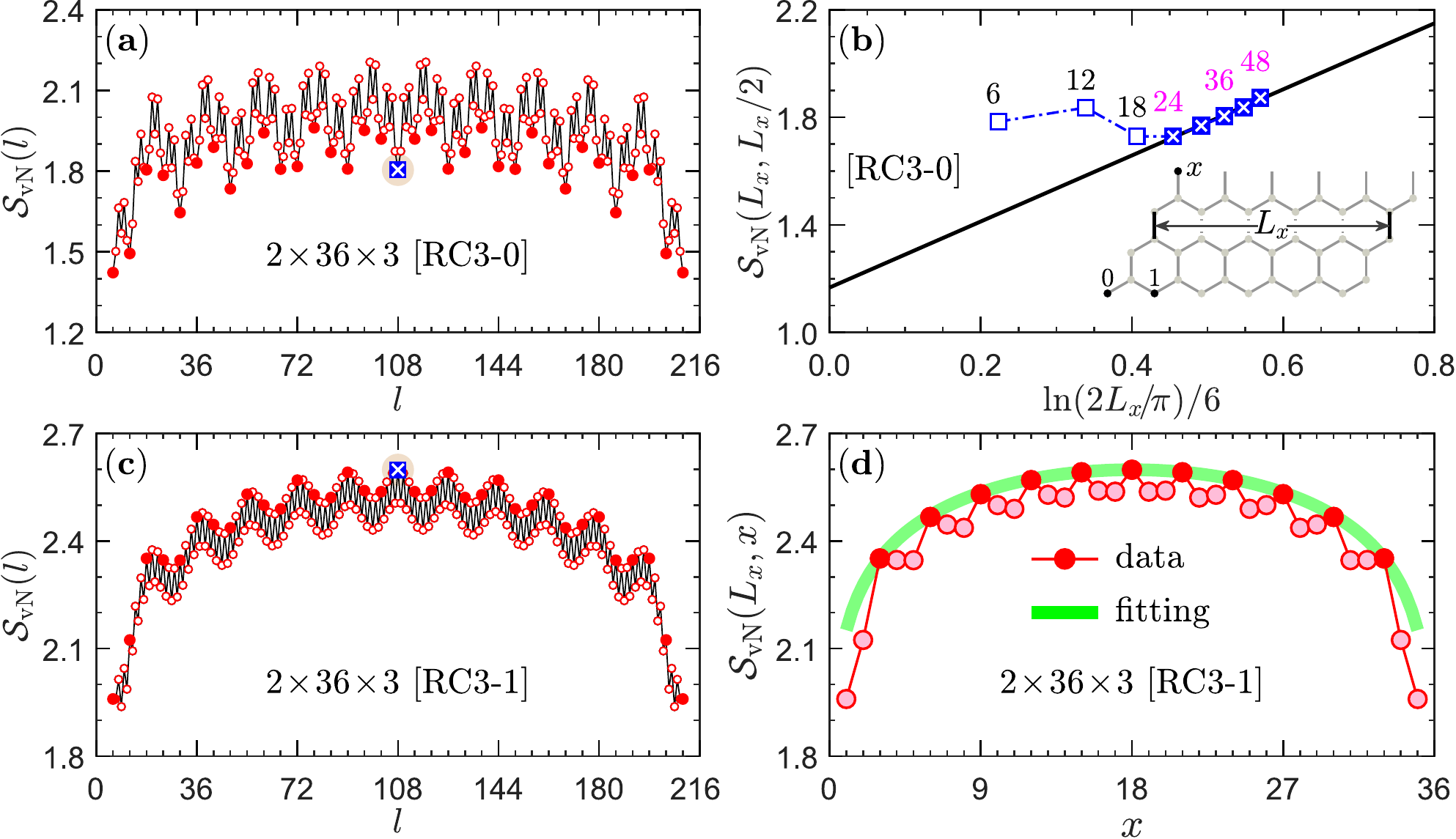}\\
\caption{The entanglement entropy and its scaling on different three-leg cylinders at $\Gamma'/\Gamma$ = 0.15 in the $\Gamma$-$\Gamma'$ model.
  (a) Entanglement entropy $\mathcal{S}_{\rm vN}(l)$ of a consecutive segment of length $l$ on a $2\times36\times3$ cylinder with the connection of RC3-0.
  The solid symbols represent the neat edge-cutting with $l$ being a multiply of 6 (i.e., the number of the sites along each column).
  The bipartite entanglement entropy with $l = N/2$ is marked as a blue square.
  (b) The bipartite entanglement entropy $\mathcal{S}_{\textrm{vN}}(L_x, L_x/2)$ under a series of $2\times L_x\times 3$ cylinders with the connection of RC3-0.
  The central charge is fitted as $1.2$ based on the results at $L_x$ = 24, 30, 36, 42, and 48.
  The inset shows the geometry of the three-leg cylinders. Depending on the way to connect the upper and lower boundaries,
  the cylinders are called as RC3-0 and RC3-1, respectively, when connecting the site $x$ (upper) to site $0$ and site $1$ (lower).
  (c) The same as (a) but for the connection of RC3-1.
  (d) Extracting the central charge $c$ from the entanglement entropy $\mathcal{S}_{\textrm{vN}}(L_x, x)$  on a $2\times36\times3$ cylinder with the connection of RC3-1.
  The fitting formula is Eq.~\eqref{EQ:EESBranch} and the central charge is estimated as 1.1.
  }\label{FIG-ChiVNE}
\end{figure*}

To gain insight into the chirality relations, we denote the ground-state doublet as $\{\ket{\textbf{K}_1},\ket{\textbf{K}_2}\}$,
with ${\bf{Q}} = \textbf{K}_{1,2}$ being the total momentum of the ground states [see inset of Fig.~\ref{FIG-GGpOP}(b)].
Since the composite operator $\hat{\Theta} = \hat{\mathcal{T}}\hat{\mathcal{P}}$ relates $\ket{\textbf{K}_{1,2}}$ to itself,
we find a staggered relation between inversion-related chiral operators in each ground state
\begin{equation}\label{eq:stagrelation1}
  \braket{ \textbf{K}_{1,2} |\hat{\chi}^{A/B}| \textbf{K}_{1,2} } = -\braket{ \textbf{K}_{1,2} |\hat{\chi}^{B/A}| \textbf{K}_{1,2} }.
\end{equation}
On the other hand, since $\hat{C}_{2b}$ relates the different ground states, we have a staggered relation between $\hat{C}_2$-related chiral operators in opposite ground states
\begin{equation}\label{eq:stagrelation2}
  \braket{ \textbf{K}_1 |\hat{\chi}^{A/B}| \textbf{K}_1 } = -\braket{ \textbf{K}_2 |\hat{\chi}^{A/B}| \textbf{K}_2 }.
\end{equation}
Note that none of the symmetries can interchange the $n$- and $p$-type chiral operators, so generically their expectation values will differ.
Taken together, these constraints account for the observed structures of the chirality in the degenerate subspace.

\subsection{Entropy and central charge}
Previously we have inferred that the CS phase is gapless based on the behavior of the energy gap with respect to the system size.
To further affirm gapless excitations in the CS phase,
we now calculate the entanglement entropy $\mathcal{S}_{\textrm{vN}}$ on the three-leg cylinder with a length of $L_x$ along the open direction.
Although the other direction is periodic, two distinct connections termed RC3-0 and RC3-1,
which correspond to identifying site $x$ with site 0 and 1 [see the inset of Fig.~\ref{FIG-ChiVNE}(b)], are treated for the sake of central charge.
For an one-dimensional critical system, it is well established that $\mathcal{S}_{\textrm{vN}}$ obeys the formula
\begin{eqnarray}\label{EQ:EESBranch}
\mathcal{S}_{\textrm{vN}}(L_x, x) = \frac{c}{6}\ln\left[\frac{2L_x}{\pi}\sin\Big(\frac{\pi x}{L_x}\Big)\right] + c'
\end{eqnarray}
where $x$ is the length of the subsystem \cite{CalCar2004,Eisert2010}.
$c$ is the central charge and $c'$ is a non-universal constant.
If the length of the subsystem is one half of the whole system, i.e., $x = L_x/2$, then the entropy exhibits the following behavior
\begin{eqnarray}\label{EQ:EESPoint}
\mathcal{S}_{\textrm{vN}}(L_x, L_x/2) = \frac{c}{6}\ln\left(\frac{2L_x}{\pi}\right) + c'.
\end{eqnarray}

Figure~\ref{FIG-ChiVNE}(a) shows the entanglement entropy $\mathcal{S}_{\textrm{vN}}(l)$ as a function of site index $l$ ($l$ = 1, 2, 3, $\cdots$, $N$)
under the cylinder of $2\times36\times3$.
Since the connection is of RC3-0 type, the gapless excitations could not propagate smoothly along the snake-like path, making the entropy be somewhat chaotic.
We have performed up to 24 sweeps during the DMRG calculation until the entropy does not change in the first few digits.
Figure~\ref{FIG-ChiVNE}(b) shows the entanglement entropy scaling according to Eq.~\eqref{EQ:EESPoint} on a series of three-leg cylinders.
When the length of the cylinder is small, the boundary effect is significant and the entropy deviates from the scaling law.
For long enough cylinder with $L_x \ge 24$, the central charge is estimated as 1.2, which is close to 1 within the numerical precision.

By contrast, in Fig.~\ref{FIG-ChiVNE}(c) we show the entanglement entropy $\mathcal{S}_{\textrm{vN}}(l)$ under the cylinder of $2\times36\times3$ with the connection of RC3-1 type.
In this case the uppermost site $6x$ at the $x$-column is connected to the lowest site $6x+1$ at the $(x+1)$-column.
So the entanglement of the whole system is enhanced and the entanglement entropy is higher than that of Fig.~\ref{FIG-ChiVNE}(a).
Nevertheless, advantage of this connection is that the gapless excitations could propagate smoothly from one edge to another,
and thus there are two distinct branches of the entanglement entropy in the $A$/$B$ sublattices of the honeycomb lattice.
In Fig.~\ref{FIG-ChiVNE}(d) we shows the entanglement entropy scaling on the $2\times36\times3$ cylinder with the RC3-1 connection according to Eq.~\eqref{EQ:EESBranch}.
Our best fitting suggests that the central charge is approximately 1.1, which is again very close to 1.
Hence, a finite central charge of 1 on the three-leg cylinder supports the existence of gapless excitations in the CS phase.

\subsection{Dynamic structure factor and modular matrix}
In this section, we attempt to gain insight into the possible topological signature of the CS phase.
We firstly calculate the dynamical structure factor (DSF) $\mathbb{S}(\mathbf{Q},\omega)$,
which encodes the information of excitation spectrum that could be measured experimentally.
It is defined via the spatiotemporal Fourier transform of the dynamical correlation,
\begin{equation}
\mathbb{S}(\mathbf{Q},\omega)=\int\mathrm{d}t \sum\limits_{i,j}\mathbf{S}_i(t)\cdot\mathbf{S}_j(0)
\mathrm{e}^{-i\mathbf{Q}\cdot(\bm{R}_i-\bm{R}_j) t} \mathrm{e}^{-i\omega t}.
\end{equation}
The calculation is carried out on a 24-site hexagonal cluster by evaluating the continued fraction representation of the DSF.
Methodologically, by preparing a special state to feed into the Lanczos algorithm,
the tridiagonal matrix that is produced hides the coefficients of the continued fraction representation \cite{GaglianoBal1987}.
The energy resolution of $\omega$ during the DSF calculation is in steps of $0.001$ and results in very sharp peaks for the finite system \cite{LaurellOkam2020}.
To plot the evolution of the DSF along the path $\boldsymbol{\Gamma}$-$\textbf{X}$-$\boldsymbol{\Gamma'}$-$\textbf{X}'$-$\boldsymbol{\Gamma}$ in the reciprocal space,
we smeared in the energies $\omega$ by convolution of the results with a Gaussian distribution of width $0.1$.
The partial integrations of the DSF are also smeared by a Gaussian distribution but with a caveat,
namely the finite cluster has only specific allowed momenta $\mathbf{Q}$ where the DSF is calculated.
After evaluating on these allowed momenta, a Gaussian distribution is used to smear out the result for intermediate $\mathbf{Q}$ values.

\begin{figure}[!ht]
\centering
    \begin{overpic}[width=0.95\columnwidth, clip]{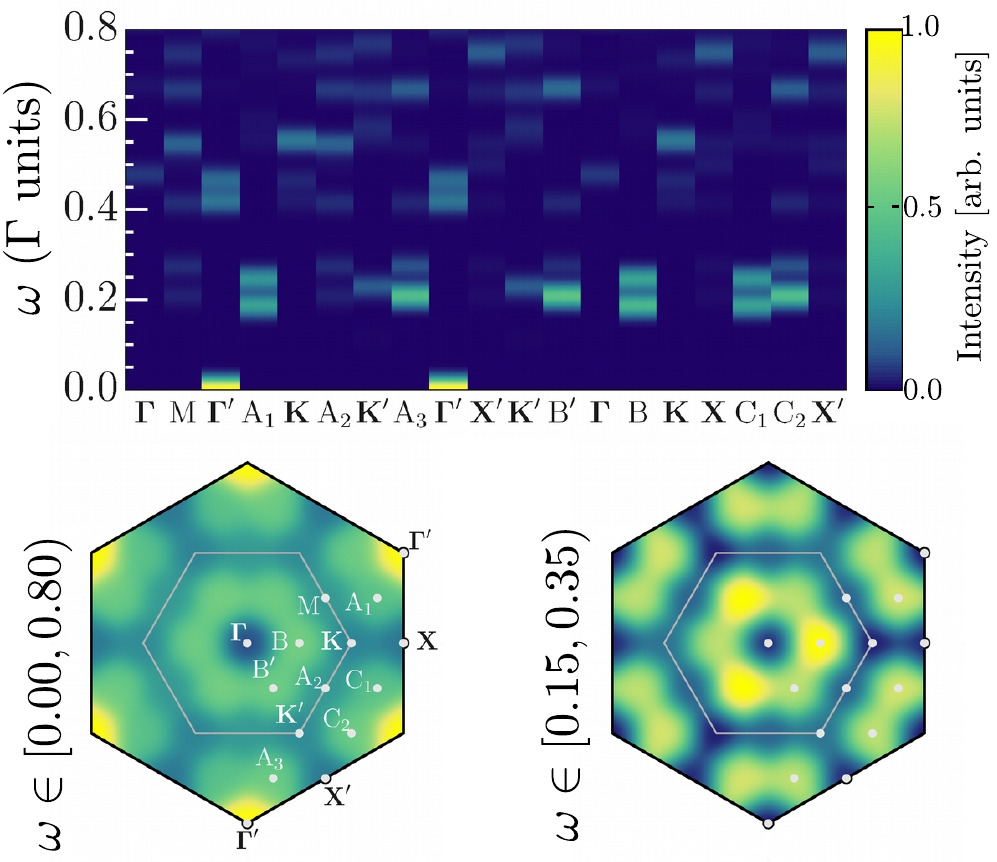}
    \put(13.5,78.0){\large{\textcolor{white}{(\textbf{a})}}}
    \put(01.0,36.0){\large{(\textbf{b})}}
    \put(53.0,36.0){\large{(\textbf{c})}}
  \end{overpic}
\caption{The DSF at $\Gamma'/\Gamma$ = 0.15 in the $\Gamma$-$\Gamma'$ model.
    (a) Spectrum of the DSF $\mathbb{S}(\textbf{Q}, \omega)$ at the high-symmetry points indicated in the left-bottom panel.
    (b) and (c) show the integrated $\mathbb{S}(\textbf{Q}, \omega)$ with respect to $\omega$ over the region of [0.0, 0.8] and [0.15, 0.35], respectively.
    }\label{FIG-DSF015}
\end{figure}

We proceed in this way for the representative point at $\Gamma'$ = 0.15 in the $\Gamma$-$\Gamma'$ model.
As can be seen from Fig.~\ref{FIG-DSF015}(a), there is a broad continuous feature in the low frequency region reminiscent of a QSL
despite the peak at $\boldsymbol{\Gamma'}$ point.
This phenomenon is in contrast to the neighboring AFM$_c$ phase whose DSF is discrete.
Figure~\ref{FIG-DSF015}(b) and (c) show the integrated DSF intensities for two different energy windows of
$\omega\in[0.0, 0.8]$ and $\omega\in[0.15, 0.35]$, respectively.
The former is an analogy of the SSF where a peak at $\boldsymbol{\Gamma'}$ point is spotted.
For the latter case, the DSF is nonsymmetric under inversion and the $C_6$ rotational symmetry
is reduced to $C_3$ due to TRS breaking \cite{HalimehPunk2016}.
Generally, the TRS breaking could manifest itself in a $\textbf{Q} \leftrightarrow -\textbf{Q}$ asymmetry in the DSF.

In the end, we check for the topological order by calculating the modular $\mathcal{S}$ matrix in the CS phase \cite{ZhangTEE2012}.
Based on the central charge being 1 and a degenerate ground state of dimension 2, semion topological order is possible \cite{rowellMTC2009}.
On each non-trivial cycle of the torus we find the minimally entangled states own a two-fold degeneracy.
In all cases we find the minimally entangled states on each cycle are nearly the same, leading to an identity matrix for the modular $\mathcal{S}$ matrix.
This result is at odd with the assumption of a CS phase with topological order, but is in line with a TRS-breaking CS ordering with short-range entanglement.
However, we cannot make definitive statement for this issue so far because of the small cluster used in our numerical calculations.
Further details can be found in the SM \cite{SuppMat}.

\section{Conclusion}
In this paper we have studied a $S = 1/2$ $JK\Gamma\Gamma'$ model with dominant $\Gamma$ interaction on a honeycomb lattice,
and in a wide proximate regime we find a novel CS phase which owns a staggered scalar chirality on the two sublattices of the geometry
and possesses a doubly degenerate ground state signifying TRS breaking.
In this CS phase, its lowest energy gap has a tendency to close, consistent with entanglement entropy scalings on three-leg cylinders where a central charge around 1 is fitted.
Further, the vanishing magnetization in large enough system size indicates that the CS phase is a magnetically disordered state,
in striking contrast to a classical CS ordering that possesses a long-range magnetic order.
To unveil the nature of the CS phase, we construct a minimally entangled state out of the degenerate doublet and calculate the modular $\mathcal{S}$ matrix,
which turns out to be approximately an identity matrix.
While we fail to probe the trail of the topological order on the small cluster,
we cannot rule out the possibility that the CS phase is a gapless QSL, in which case the modular matrix is not well-defined.
This scenario is supported by the DSF calculation where a broad continuous feature in the low frequency region is observed.
We hope that our discovery will stimulate further theoretical studies of the alluring CS phase.


\begin{acknowledgments}
We thank S.-S. Gong and Y.-B. Kim for helpful discussion.
Q.L. also appreciates X. Wang and J. Zhao for the collaboration on a related topic.
This research was supported by the NSERC Discovery Grant No. 06089-2016,
the Centre for Quantum Materials at the University of Toronto
and the Canadian Institute for Advanced Research.
H.-Y.K. also acknowledged funding from the Canada Research Chairs Program.
Computations were performed on the GPC and Niagara supercomputers at the SciNet HPC Consortium.
SciNet is funded by: the Canada Foundation for Innovation under the auspices of Compute Canada;
the Government of Ontario; Ontario Research Fund - Research Excellence; and the University of Toronto.
\end{acknowledgments}

%




\clearpage

\onecolumngrid

\newpage

\newcounter{equationSM}
\newcounter{figureSM}
\newcounter{tableSM}
\stepcounter{equationSM}
\setcounter{section}{0}
\setcounter{equation}{0}
\setcounter{figure}{0}
\setcounter{table}{0}
\setcounter{page}{1}
\makeatletter
\renewcommand{\theequation}{\textsc{S}\arabic{equation}}
\renewcommand{\thefigure}{\textsc{S}\arabic{figure}}
\renewcommand{\thetable}{\textsc{S}\arabic{table}}


\begin{center}
{\large{\bf Supplemental Material for\\
 ``Spontaneous Chiral-Spin Ordering in Spin-Orbit Coupled Honeycomb Magnets''}}
\end{center}
\begin{center}
Qiang Luo$^{1}$, P. Peter Stavropoulos$^{1}$, Jacob S. Gordon$^{1}$, and Hae-Young Kee$^{1,\,2}$\\
\quad\\
$^1$\textit{Department of Physics, University of Toronto, Toronto, Ontario M5S 1A7, Canada}\\
$^2$\textit{Canadian Institute for Advanced Research, Toronto, Ontario, M5G 1Z8, Canada}\\
(Dated: January 2, 2022)
\quad\\
\end{center}

\twocolumngrid


In this Supplemental Material (SM), we present additional results that support the main findings of the main text.
The structure of the SM is as follows.
In Sec.~\textcolor{red}{I}, we use the parallel tempering Monte Carlo (PTMC) method  \cite{SMMetropolis1953,SMHukushimaNemoto1996}
to study the classical phase diagram (see Fig.~\ref{FIG-JKGGpCPD})
of the $JK\Gamma\Gamma'$ model with a dominating $\Gamma$ interaction.
Our main focus is the large-unit-cell (LUC) magnetically ordered states.
In the following Sec.~\textcolor{red}{II}, we reinforce the properties of scalar spin chirality and the energy gap is the chiral-spin (CS) ordering.
Finally, in Sec.~\textcolor{red}{III} we present the momentum-resolved ED calculation and show the details of modular $\mathcal{S}$-matrix.

\section{Classical phase diagrams of the $JK\Gamma\Gamma'$ model}\label{SMSEC:PTMC}

\begin{figure}[!ht]
\centering
\includegraphics[width=0.95\columnwidth, clip]{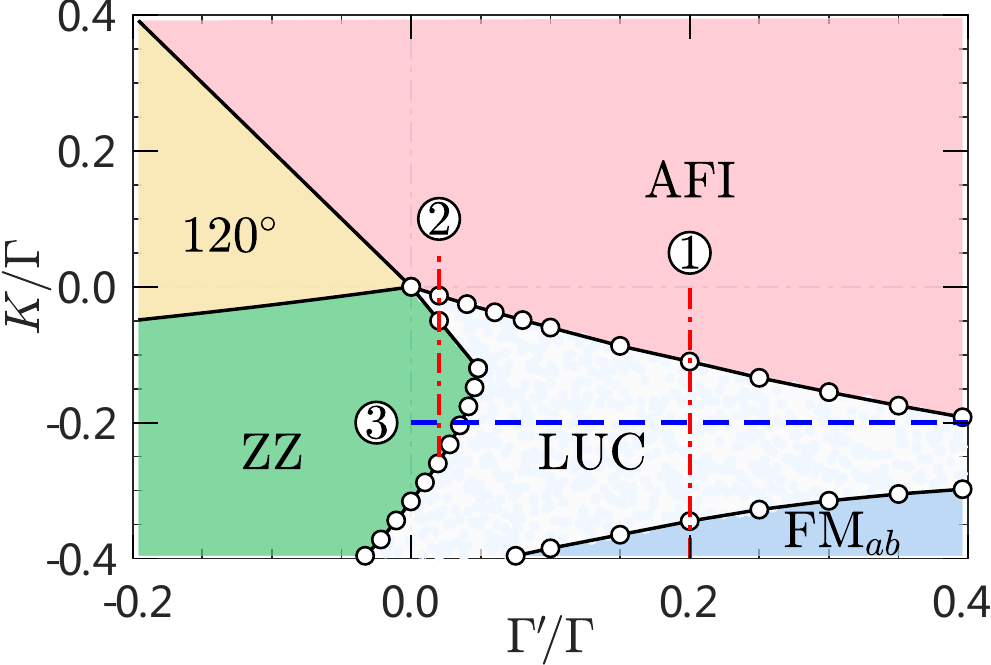}\\
\caption{Classical phase diagram of the $JK\Gamma\Gamma'$ model with the dominant AFM $\Gamma$ interaction.
    Here, $\Gamma$ = 1 and $J = K$. There are four conventional magnetically ordered states,
    AFM$_c$ phase, 120$^{\circ}$ phase, zigzag phase, and FM$_{ab}$ phase,
    and also a LUC region with dozens of LUC phases and some interim incommensurate phases.
    The line $\textcircled{1}$ ($\Gamma' = 0.20$), line $\textcircled{2}$ ($\Gamma' = 0.02$), and line $\textcircled{3}$ ($K = -0.20$)
    are the selected cuttings in the LUC region.}
    \label{FIG-JKGGpCPD}
\end{figure}

We begin by studying the classical phase diagram of the generic $JK\Gamma\Gamma'$ model
whose Hamiltonian is shown in Eq.~(\textcolor{red}{8}) in the main text.
We take $\Gamma = 1$ as the energy unit and consider the diagonal interactions on equal footing,
namely, \textcolor{blue}{$\boldsymbol{J = K}$}.
The classical phase diagram shown in Fig.~\ref{FIG-JKGGpCPD} is mapped out by PTMC method.
It includes the AFM$_c$ phase (pink), 120$^{\circ}$ phase (yellow), zigzag phase (green), FM$_{ab}$ phase (blue),
and also dozens of LUC phases and some incommensurate phases (shown in the white region).

\begin{figure}[!ht]
\centering
\includegraphics[width=0.90\columnwidth, clip]{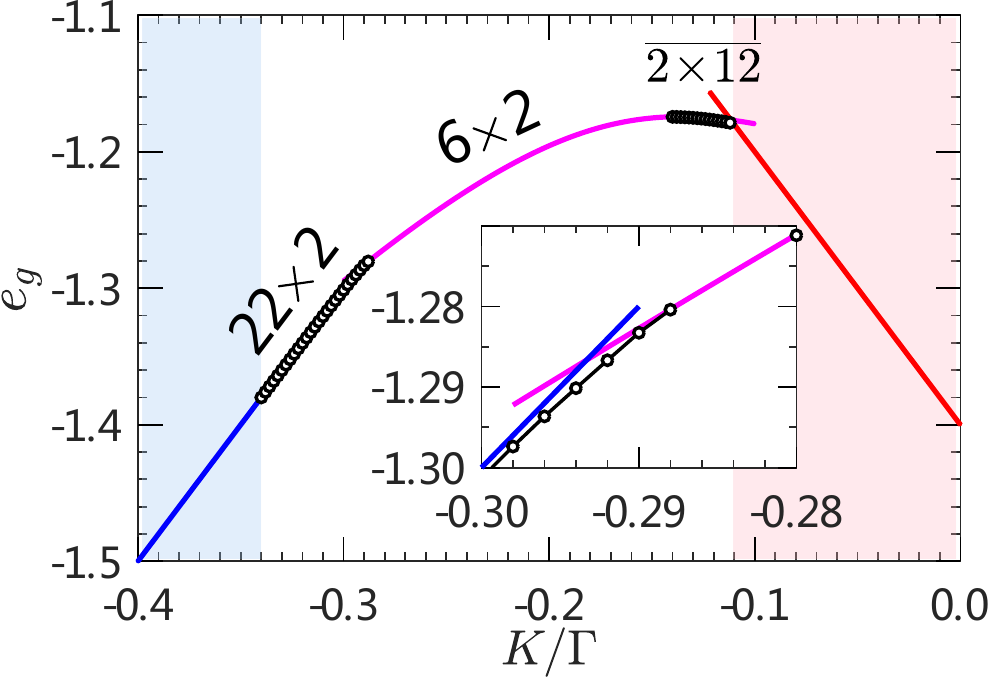}\\
\caption{Classical ground state energy $e_g$ along the line $\textcircled{1}$ of $\Gamma' = 0.20$, as depicted in Fig.~\ref{FIG-JKGGpCPD}.
    The blue region is the FM$_{ab}$ phase while the pink region is the AFM$_c$ phase.
    There are $22\times2$ phase (i.e., 44-site phase), $6\times2$ phase (i.e., 12-site phase), and $\overline{2\times12}$ phase (i.e., 24-site phase) in the LUC region.
    Inset: zoom-in of the energy near the $22\times2$ phase.}
    \label{FIG-Gp020CPD}
\end{figure}

In the following, we use the PTMC method to clarify how do the LUC phases look like.
The simulations are performed on XC cylinders of $L_x\times L_y$ (cf. Fig.~\ref{FIG-Gp020OF2x6})
where $L_x$~($L_y$) is the number of sites along the $\textbf{Z}$~($\textbf{X/Y}$)-bond.
To begin with, we consider the transitions along the \textcolor{red}{line $\textcircled{1}$} of $\Gamma' = 0.20$,
which crosses the very middle of the LUC region.
Apart from the FM$_{ab}$ and AFM$_c$ phases at the boundaries,
three LUC phases, $22\times2$, $6\times2$, and $\overline{2\times12}$, are identified (see Fig.~\ref{FIG-Gp020CPD}).
Here, $p\times q$ denotes the number of sites along the $L_x$ and $L_y$ directions within a unit cell,
and the over line stands for a noncoplanar phase.

\begin{figure}[!ht]
\centering
\includegraphics[width=0.90\columnwidth, clip]{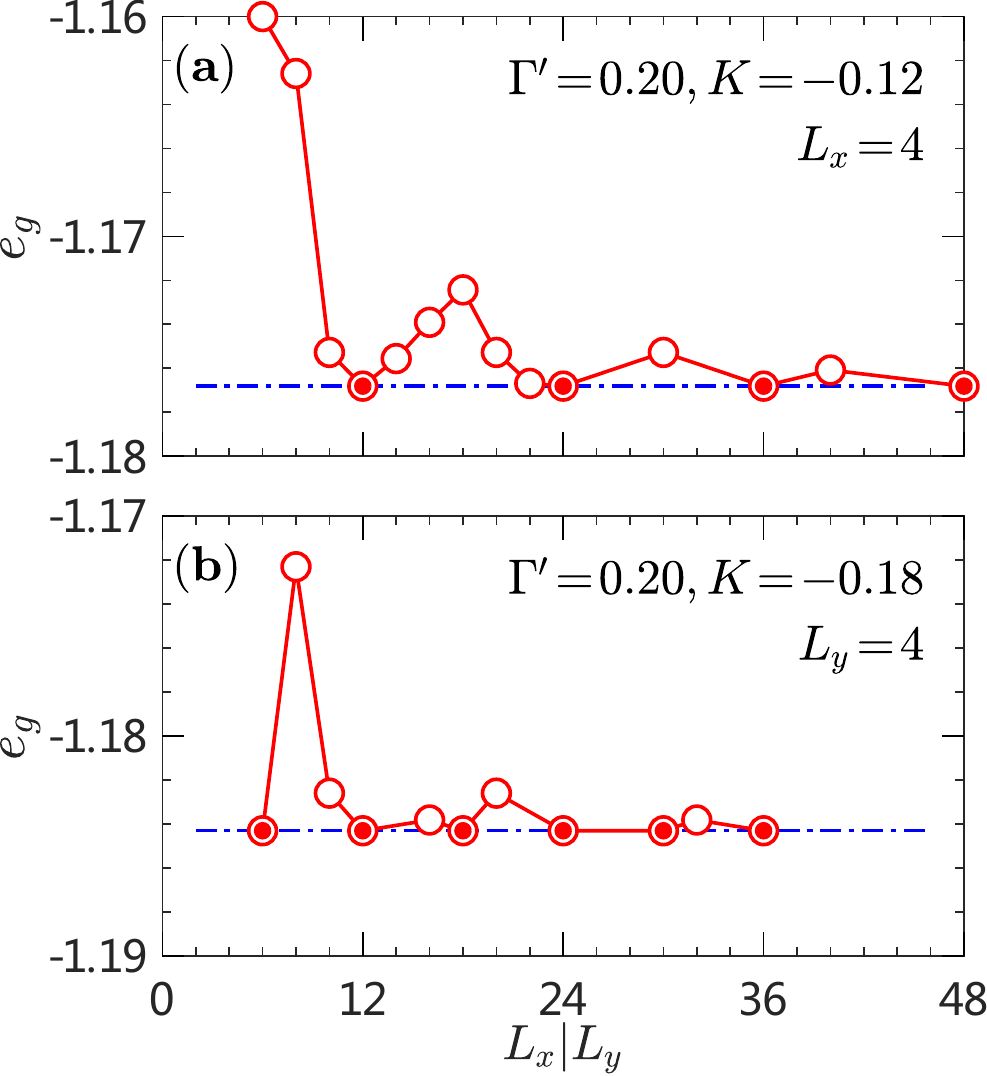}\\
\caption{Classical ground-state energy for the restricted cluster of $L_x\times L_y$.
    (a) For $(\Gamma', K) = (0.20, -0.12)$, $L_x =4$ and the energy has a minimal when $L_y = 12, 24, 36$, and 48.
    (b) For $(\Gamma', K) = (0.20, -0.18)$, $L_y =4$ and the energy has a minimal when $L_x = 6, 12, 18, 24, 30$, and 36.}
    \label{FIG-Gp020CUC}
\end{figure}

\begin{figure}[!ht]
\centering
\includegraphics[width=0.90\columnwidth, clip]{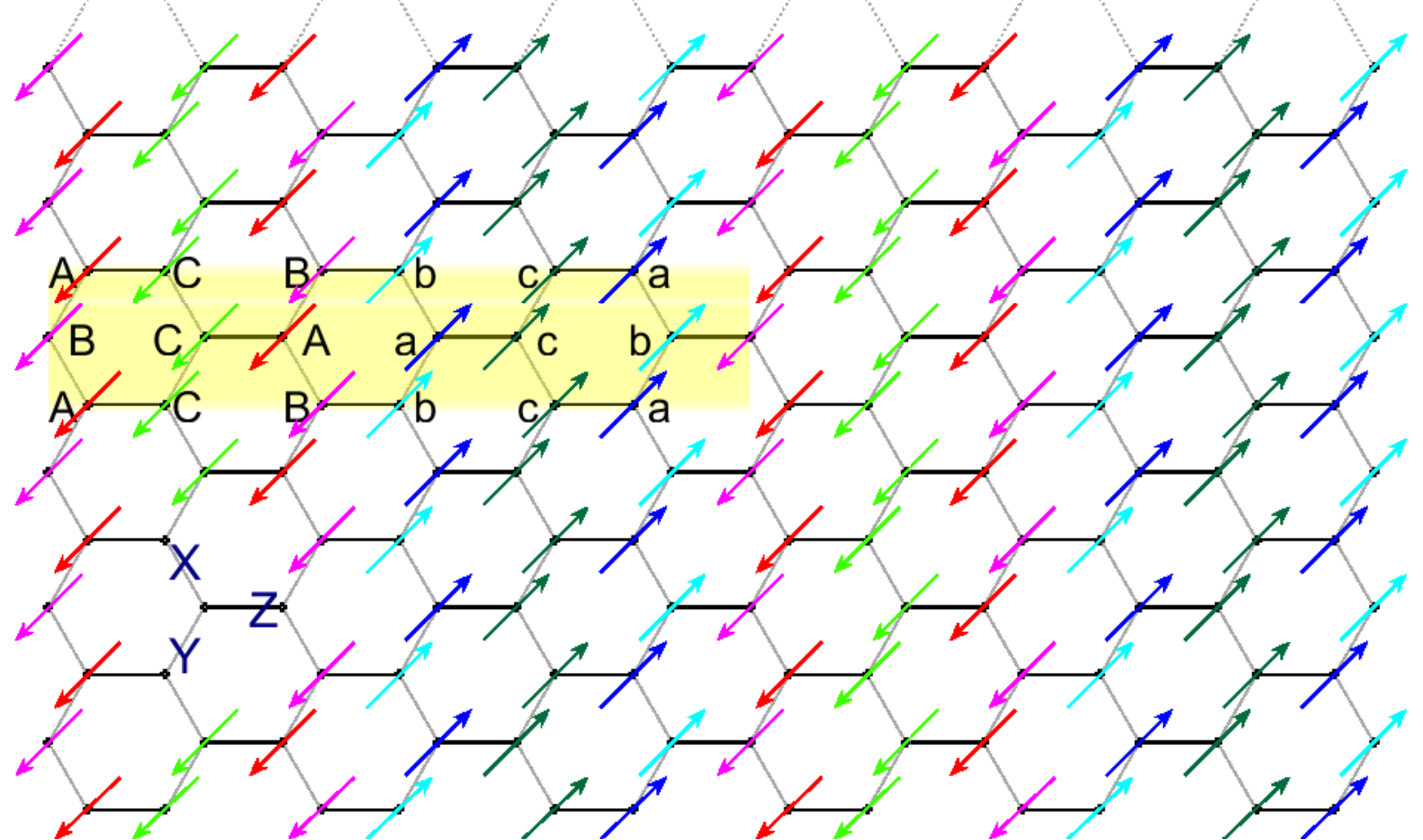}\\
\caption{Spin pattern of the $6\times2$ phase.
    The spins are parameterized by angles $\theta_i$ and $\phi_i$ and the arrows are plotted by $\phi_i$.
    The shaded region is the unit cell which contains six different kinds of spins.
    $\hat{\textbf{S}}_A/\hat{\textbf{S}}_a$, $\hat{\textbf{S}}_B/\hat{\textbf{S}}_b$,
    and $\hat{\textbf{S}}_C/\hat{\textbf{S}}_c$ are antiparallel, respectively.}
    \label{FIG-Gp020OF2x6}
\end{figure}

The unit cell is determined by a careful inspection of the spin patterns.
To precisely determine the minimal unit cell of each phase, we adopt a series of clusters of different $L_x$ and $L_y$.
Usually, one of the two ($L_x$ and $L_y$) is 2, while the other is sensitive to the interaction parameters.
For the parameters $(\Gamma', K) = (0.20, -0.12)$,
we fix $L_x = 4$ (so as to apply a periodic boundary condition (PBC) properly) and tune $L_y$ from 6 to 48.
We find that the energy have a minimal when $L_y$ = 12, 24, 36, and 48, showing a 12-site periodicity (see Fig.~\ref{FIG-Gp020CUC}(a)).
Therefore, it is identified as a $\overline{2\times12}$ phase.
Similarly, for the parameters $(\Gamma', K) = (0.20, -0.18)$,
we fix $L_y = 4$ and tune $L_x$ from 6 to 36.
We find that the energy have a minimal when $L_y$ is the multiply of six,
revealing a $6\times2$ phase (see Fig.~\ref{FIG-Gp020CUC}(b)).
We note that the $\overline{2\times12}$ phase is a noncoplanar phase and the spin patterns are complicated.
However, the $6\times2$ phase is a coplanar phase whose energy only depends on three undetermined angles.

To be specific, the typical spin patterns of $6\times2$ phase is shown in Fig.~\ref{FIG-Gp020OF2x6}.
Let us parameterize the classical spins as
$\hat{\textbf{S}}_i(\theta_i, \phi_i) = S(\sin\theta_i\cos\phi_i, \sin\theta_i\sin\phi_i, \cos\theta_i)$,
then we have
$S_A\big(\theta_1, \frac{5\pi}{4}\big)$,
$S_B\big(\theta_2, \frac{5\pi}{4}\big)$, and
$S_C\big(\theta_3, \frac{5\pi}{4}\big)$.
Since $S_{A/B/C}$ and $S_{a/b/c}$ are antiparallel, we then have
$S_a\big(\pi-\theta_1, \frac{\pi}{4}\big)$,
$S_b\big(\pi-\theta_2, \frac{\pi}{4}\big)$, and
$S_c\big(\pi-\theta_3, \frac{\pi}{4}\big)$.
Suppose that $\mathcal{H}_{PQ}^{\gamma}$ is the $\gamma$-bond energy between spin $S_P$ and $S_Q$,
we find the form of the ground-state energy
\begin{equation*} 
e_g = \frac{1}{6}\Big[2(\mathcal{H}_{AB}^x \!+\! \mathcal{H}_{AB}^y \!+\! \mathcal{H}_{AC}^z)
        \!+\! (\mathcal{H}_{CC}^x \!+\! \mathcal{H}_{CC}^y \!-\! \mathcal{H}_{BB}^z)\Big].
\end{equation*}
With the help of symbol calculation in \textsc{Matlab} or \textsc{Mathematica},
it is possible to have an analytical expression of the energy,
which only depends on three parameters ($\theta_1$, $\theta_2$, and $\theta_3$).
Furthermore, the ground-state energy $e_g$ could be determined by \textsf{fminsearch} function.
For example, when $(\Gamma', K) = (0.20, -0.18)$ we find that
$\big(\theta_1, \theta_2, \theta_3\big) = \big(1.9468, 0.5234, 0.2472\big)\pi$
and $e_g = -1.18431605$, which is fairly consistent with the one shown in Fig.~\ref{FIG-Gp020CUC}(b).

\begin{figure}[!ht]
\centering
\includegraphics[width=0.90\columnwidth, clip]{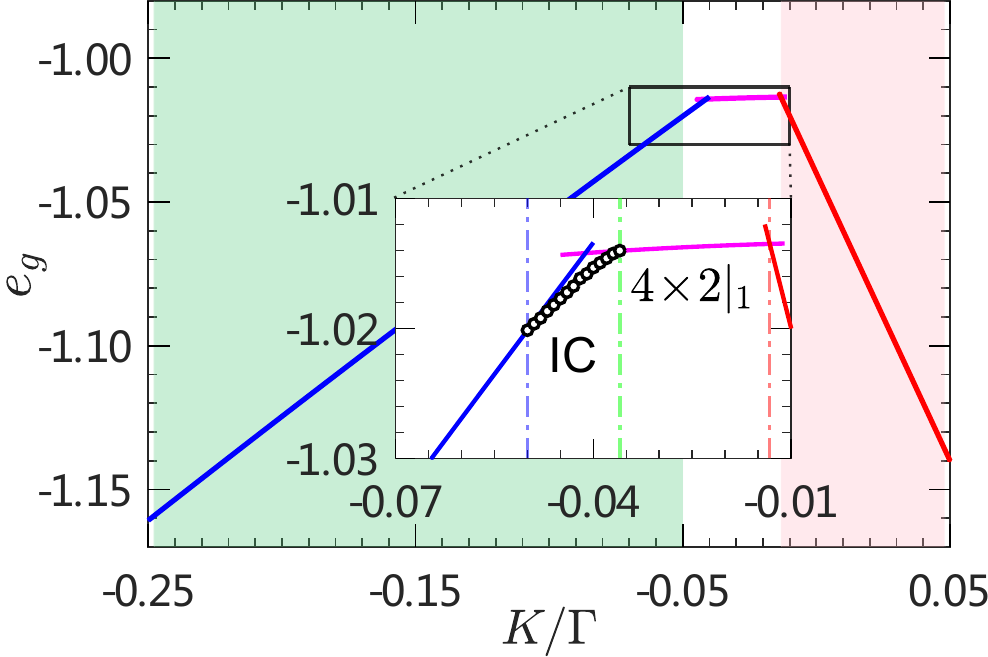}\\
\caption{Classical ground state energy $e_g$ along the line $\textcircled{2}$ of $\Gamma' = 0.02$, as depicted in Fig.~\ref{FIG-JKGGpCPD}.
    The green region is the zigzag phase while the pink region is the AFM$_c$ phase.
    A $4\times2|_1$ phase and an incommensurate (IC) phase are found when $-0.050 < K < -0.013$.}
    \label{FIG-Gp002CPD}
\end{figure}

We then shift to the \textcolor{red}{line $\textcircled{2}$} of $\Gamma' = 0.02$,
which is very close to the AFM $\Gamma$ limit.
As shown in Fig.~\ref{FIG-Gp002CPD},
near the AFM$_c$ side there is a $4\times2|_1$ phase whose energy is not sensitive to $J, K$.
The subscript here represents one sort of the spin pattern within the unit cell,
as other kinds of spin arrangements are also possible.
With the increase of the FM Kitaev interaction,
an incommensurate phase appears when $-0.050 < K < -0.036$ (see inset).
Physically, since classical moment directions of two neighboring magnetically ordered phases are not the same,
there could be an incommensurate phase to intervene them.

\begin{figure}[!ht]
\centering
\includegraphics[width=0.90\columnwidth, clip]{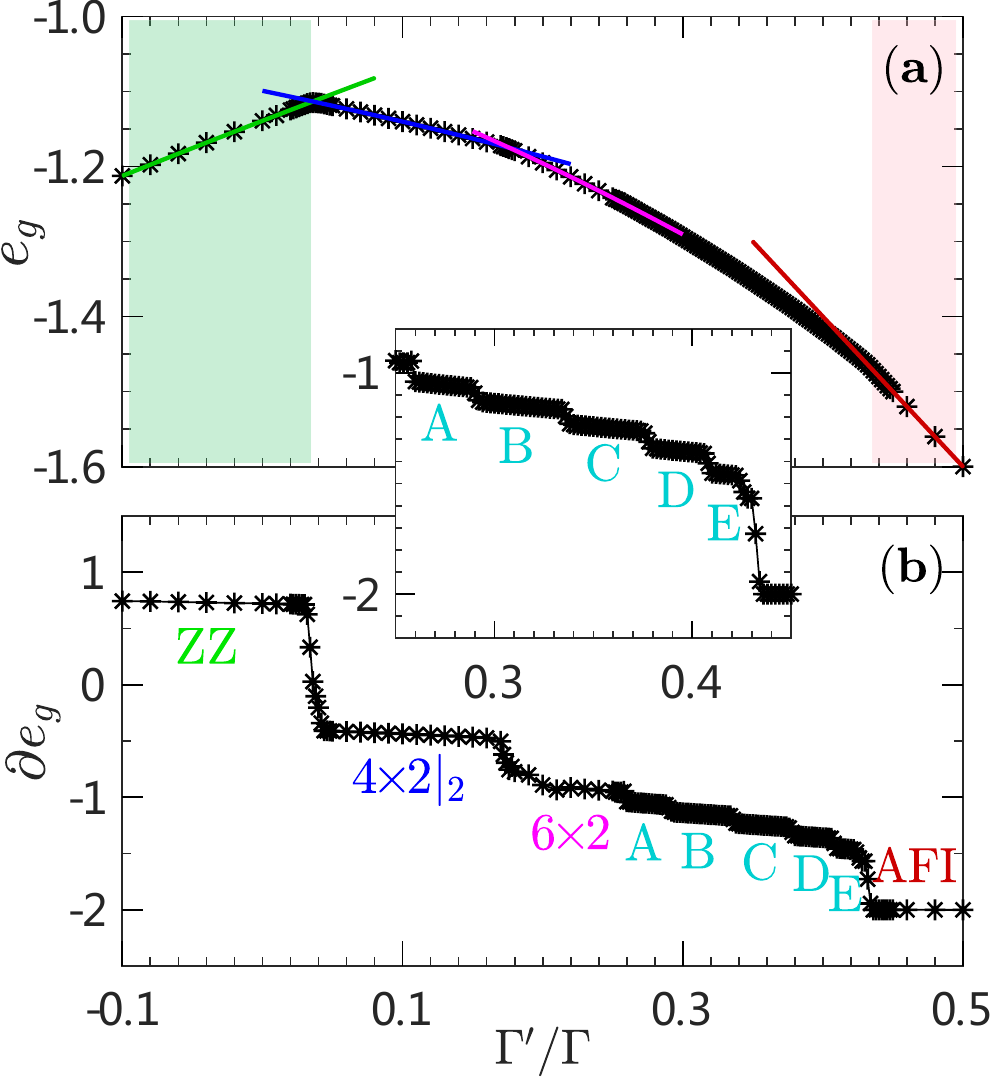}\\
\caption{(a) Classical ground state energy $e_g$ and (b) its energy derivative $\partial e_g$
    along the line $\textcircled{3}$ of $K = -0.20$, as depicted in Fig.~\ref{FIG-JKGGpCPD}.
    The green region is the zigzag phase while the pink region is the AFM$_c$ phase.
    There are seven distinct LUC phases between the zigzag phase and the AFM$_c$ phase.
    These LUC phases are known to be the $4\times2|_2$ phase, the $6\times2$ phase, and the (A, B, C, D, E) phases.
    Details of them are shown in the text.}
    \label{FIG-JK020CPD}
\end{figure}

Although there are only a few LUC phases to meet when $\Gamma'$ is fixed,
things are dramatically different when tuning $\Gamma'$ interaction along the horizontal lines.
We focus on the \textcolor{blue}{line $\textcircled{3}$} of $K = -0.20$,
which crosses the phase diagram from the zigzag phase to the AFM$_c$ phase.
The classical energy $e_g$ are shown in Fig.~\ref{FIG-JK020CPD}(a), where $\Gamma'$ is in the range of $-0.1$ to 0.5.
Near the zigzag phase, we find a $4\times2|_2$ phase,
which has the same size of the unit cell as the $4\times2|_1$ phase, but the spins are arranged differently.
Furthermore, the $6\times2$ phase in Fig.~\ref{FIG-Gp020CPD} is found alongside.
We also plot the analytical classical energy $e_g$ of the $4\times2|_2$~(blue) and $6\times2$~(pink) phases,
which matches with the PTMC result well.
To further recognize the LUC phases, we plot the energy derivative, as shown in Fig.~\ref{FIG-JK020CPD}(b).
We stress that the LUC phases are very robust once we find the proper unit cells associating with the underlying magnetic structures.
There are many jumps, which are signal of phase transitions.
The plateau between any two jumps are recognized as a phase.
The plateaus around $\Gamma' = 0.10$ and 0.20 are known to be the $4\times2|_2$ phase and $6\times2$ phase, respectively.
Five extra new LUC phases dubbed A, B, C, D, and E, appear when $0.26 < \Gamma' < 0.43$, see inset.
These phases are known to be the $20\times2$ phase, $\overline{2\times12}$ phase, $8\times2$ phase,
$10\times2$ phase, and $12\times2$ phase, respectively.
We emphasize that the $\overline{2\times12}$ phase is the same as that in Fig.~\ref{FIG-Gp020CPD},
which is known to be a noncoplanar phase with a finite scalar chirality.

\section{The Scalar Chirality Phase}\label{SMSEC:SSC}

\subsection{Basic relation of the staggered scalar chirality}
\begin{figure}[!ht]
\centering
\includegraphics[width=0.85\columnwidth, clip]{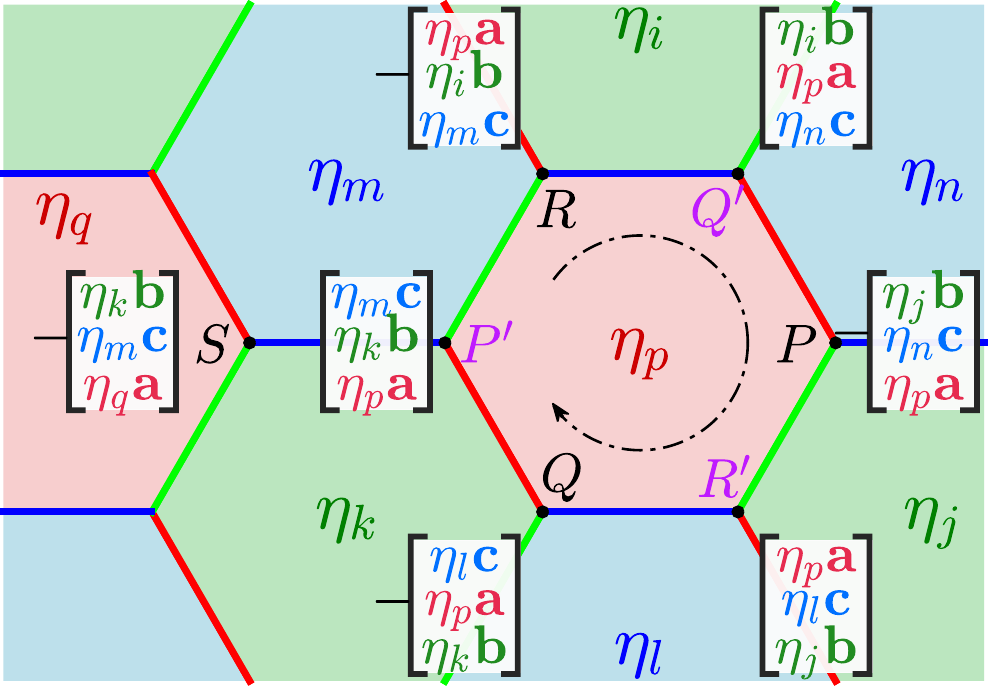}\\
\caption{Classical spin components around a given hexagonal plaquette with an Ising variable $\eta_p$ or $\eta_q$ (red).
    Plaquettes with Ising variables ($\eta_i$, $\eta_j$, $\eta_k$) (green) and ($\eta_l$, $\eta_m$, $\eta_n$) (blue)
    represent the other two interpenetrating triangular sublattices.
    Here, sites at ($P$, $Q$, $R$, $S$) belong to the $A$ sublattice,
    while ($P'$, $Q'$, $R'$) are located at the $B$ sublattice.}
    \label{FIGSM-HexChi}
\end{figure}

One of the peculiar properties of the scalar chirality is that it is inherently \textit{staggered} within each hexagonal plaquette
for the degenerate ground state of the classical AFM $\Gamma$ model.
As proposed in Ref.~ \cite{SMRousPerk2017},
the classical spin $\hat{\textbf{S}}_n$ could be parameterized by
$\hat{\textbf{S}}_n = (\eta_i a,\eta_j b,\eta_k c)$
where ($a$, $b$, $c$) = ($|S^x_n|$, $|S^y_n|$, $|S^z_n|$)
and $\eta_p = \pm 1$ is an Ising variable, see Fig.~\ref{FIGSM-HexChi}.
By definition, for spins at ($P$, $Q$, $R$) of the $A$ sublattice we have
\begin{align}\label{SMEQ:ChiA}
\chi_{PQR}
&= \hat{\textbf{S}}_P \cdot (\hat{\textbf{S}}_Q \times \hat{\textbf{S}}_R) =
\left|\begin{array}{cccc}
    S_P^x    &    S_P^y   & S_P^z \\
    S_Q^x    &    S_Q^y   & S_Q^z \\
    S_R^x    &    S_R^y   & S_R^z \\
\end{array}\right|  \nonumber\\
&= S_P^x\big(S_Q^yS_R^z - S_Q^zS_R^y\big)
 + S_P^y\big(S_Q^zS_R^x - S_Q^xS_R^z\big)    \nonumber\\
&\quad\, + S_P^z\big(S_Q^xS_R^y - S_Q^yS_R^x\big) \nonumber\\
&= -\big\{\eta_jb (\eta_pa\eta_mc \!-\! \eta_kb\eta_ib) + \eta_nc (\eta_kb\eta_pa \!-\! \eta_lc\eta_mc) \nonumber\\
&\qquad\;\; + \eta_pa (\eta_lc\eta_ib \!-\! \eta_pa\eta_pa)\big\}   \nonumber\\
&= \eta_p^3a^3 + \eta_i\eta_j\eta_kb^3 + \eta_l\eta_m\eta_nc^3  \nonumber\\
&\quad\, -abc\eta_p\big(\eta_i\eta_l + \eta_j\eta_m + \eta_k\eta_n\big).
\end{align}

On the other hand, we find the scalar chirality for spins at ($P'$, $Q'$, $R'$) of the $B$ sublattice is
\begin{align}\label{SMEQ:ChiB}
\chi_{P'Q'R'}
&= \hat{\textbf{S}}_{P'} \cdot (\hat{\textbf{S}}_{Q'} \times \hat{\textbf{S}}_{R'}) \nonumber\\
&= S_{P'}^x\big(S_{Q'}^yS_{R'}^z \!-\! S_{Q'}^zS_{R'}^y\big)
 \!+\! S_{P'}^y\big(S_{Q'}^zS_{R'}^x \!-\! S_{Q'}^xS_{R'}^z\big)    \nonumber\\
&\quad\, + S_{P'}^z\big(S_{Q'}^xS_{R'}^y - S_{Q'}^yS_{R'}^x\big)    \nonumber\\
&= \eta_mc (\eta_pa\eta_jb \!-\! \eta_nc\eta_lc) + \eta_kb (\eta_nc\eta_pa \!-\! \eta_ib\eta_jb) \nonumber\\
&\qquad\;\; + \eta_pa (\eta_ib\eta_lc \!-\! \eta_pa\eta_pa)\big\}   \nonumber\\
&= abc\eta_p\big(\eta_i\eta_l + \eta_j\eta_m + \eta_k\eta_n\big)    \nonumber\\
&\quad\, -\big(\eta_p^3a^3 + \eta_i\eta_j\eta_kb^3 + \eta_l\eta_m\eta_nc^3\big).
\end{align}

Combining Eq.~\eqref{SMEQ:ChiA} and Eq.~\eqref{SMEQ:ChiB}, we can immediately deduce that
\begin{equation}\label{SMEQ:ChiAmB}
\boxed{\chi_{PQR} = -\chi_{P'Q'R'}}.
\end{equation}

The scalar chirality in Eq.~\eqref{SMEQ:ChiA} and Eq.~\eqref{SMEQ:ChiB} are defined in the hexagons and thus termed as $n$-type.
If the spins are centered at a real site, we call it as $\chi_p$.
For example, spins at ($S$, $R$, $Q$) belongs to the $A$ sublattice and there is a site $P'$ in the center.
In this case, we have
\begin{align}\label{SMEQ:ChiAPtype}
\chi_{SRQ}
&= \hat{\textbf{S}}_S \cdot (\hat{\textbf{S}}_R \times \hat{\textbf{S}}_Q) \nonumber\\
&= S_S^x\big(S_R^yS_Q^z - S_R^zS_Q^y\big)
 + S_S^y\big(S_R^zS_Q^x - S_R^xS_Q^z\big)    \nonumber\\
&\quad\, + S_S^z\big(S_R^xS_Q^y - S_R^yS_Q^x\big) \nonumber\\
&= -\big\{\eta_kb (\eta_ib\eta_kb \!-\! \eta_mc\eta_pa) + \eta_mc (\eta_mc\eta_lc \!-\! \eta_pa\eta_kb) \nonumber\\
&\qquad\;\; + \eta_qa (\eta_pa\eta_pa \!-\! \eta_ib\eta_lc)\big\}   \nonumber\\
&= abc\big(2\eta_p\eta_k\eta_m + \eta_q\eta_i\eta_l\big) \nonumber\\
&\quad\, -(\eta_q\eta_p^2a^3 + \eta_i\eta_k^2b^3 + \eta_l\eta_m^2c^3)   \nonumber\\
&= abc\big(2\eta_p\eta_k\eta_m + \eta_q\eta_i\eta_l\big) -(\eta_qa^3 + \eta_ib^3 + \eta_lc^3),
\end{align}
where in the last step we use the property that $\eta^2 = 1$.

\begin{figure}[!ht]
\centering
\includegraphics[width=0.66\columnwidth, clip]{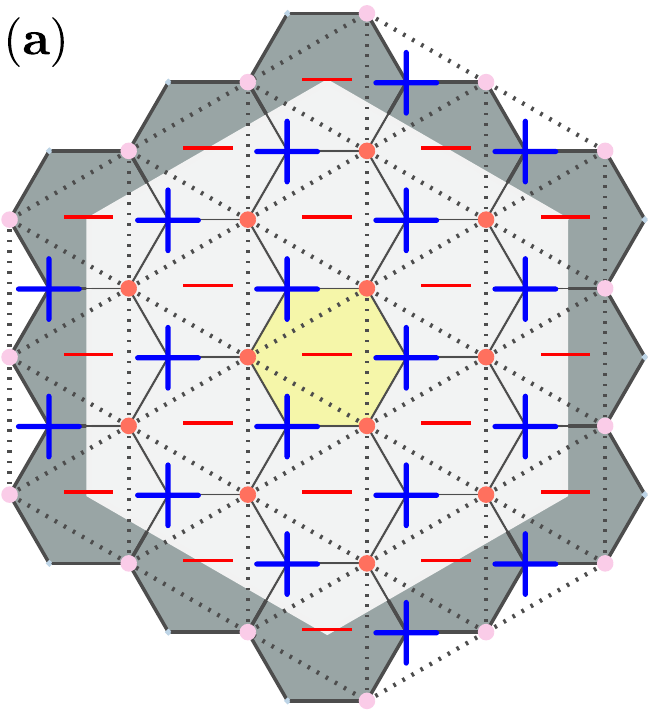}\\
\caption{Sign pattern of the scalar chirality $\hat{\chi}_{ijk}$ for the \textit{classical} magnetically ordered CS state.
    Here, only $\chi_p$ (blue) and $\chi_n$ (red) of the $B$ sublattice are shown.}
    \label{FIGSM-ChiPatternCvsQ}
\end{figure}

In what follows, we shall choose a special CS ordering where all the scalar chirality of the same kinds are uniformly distributed.
Taking $\chi_n^A$ as an example, we have
\begin{equation}\label{SMEQ:SameChi}
\cdots = \hat{\chi}_{7,19,17} = \hat{\chi}_{9,17,15} = \hat{\chi}_{11,15,13} = \cdots
\end{equation}
where the site index is shown in Fig.~\textcolor{red}{1} in the main text.
Since ($a, b, c$) are randomly chosen, equation series in Eq.~\eqref{SMEQ:SameChi} imply that
Ising variables $\eta$ in each triangular sublattice should be the same, namely,
\begin{align}\label{SMEQ:EtaABC}
\left\{
  \begin{array}{l}
    \eta_1 = \eta_4 = \eta_7 = \eta_{10} = \cdots \equiv \textcolor{red}{\eta_a}   \\
    \eta_2 = \eta_5 = \eta_8 = \eta_{11} = \cdots \equiv \textcolor{green}{\eta_b}   \\
    \eta_3 = \eta_6 = \eta_9 = \eta_{12} = \cdots \equiv \textcolor{blue}{\eta_c}
  \end{array}
\right..
\end{align}
Therefore, Eq.~\eqref{SMEQ:EtaABC} strongly restricts $\eta$-values.
Since only three free $\eta$ survive, the \textit{conventional} CS phase
stemming from the classical degenerate ground state
have $2^3 = 8$ folds degeneracy.
Furthermore, under the constraint of Eq.~\eqref{SMEQ:EtaABC},
Eq.~\eqref{SMEQ:ChiA} and Eq.~\eqref{SMEQ:ChiAPtype} reduce to
\begin{align*}
\left\{
    \begin{array}{l}
        \chi_n^{A} \equiv \chi_{P,Q,R} = \eta_aa^3 + \eta_bb^3 + \eta_cc^3 -3abc \eta_a\eta_b\eta_c      \nonumber\\
        \chi_p^{A} \equiv \chi_{S,R,Q} = 3abc \eta_a\eta_b\eta_c - (\eta_aa^3 + \eta_bb^3 + \eta_cc^3)   \nonumber\\
    \end{array}
\right.,
\end{align*}
namely,
\begin{equation}\label{SMEQ:ChiApmAn}
\boxed{\chi_n^{A} = -\chi_p^{A}}.
\end{equation}

To conclude, by virtue of Eq.~\eqref{SMEQ:ChiAmB} and Eq.~\eqref{SMEQ:ChiApmAn}, we find that
\boldsymbol{$\chi_n^{A} = -\chi_n^{B} = -\chi_p^{A} = \chi_p^{B}$}.
Equivalently, we have two following statements:
\begin{itemize}
  \item For either $\chi_p$ and $\chi_n$, there is a staggered relation between $A$ and $B$ sublattices,
    namely, $\chi_{t}^{A} = -\chi_{t}^{B}$ with $t = p, n$.
  \item Within $A$ or $B$ sublattice, $\chi_p$ and $\chi_n$ are also staggered,
    namely, $\chi_p^{\triangle} = -\chi_n^{\triangle}$ with $\triangle = A, B$.
\end{itemize}

For the benefit of visualization,
we have presented the pattern of the scalar chirality in Fig.~\ref{FIGSM-ChiPatternCvsQ}(a)
where $\chi_p$ and $\chi_n$ of the $B$ sublattice is shown.
It is clearly shown that there is a sign differences between $\chi_p$ and $\chi_n$.

\begin{figure}[!ht]
\centering
\includegraphics[width=0.90\columnwidth, clip]{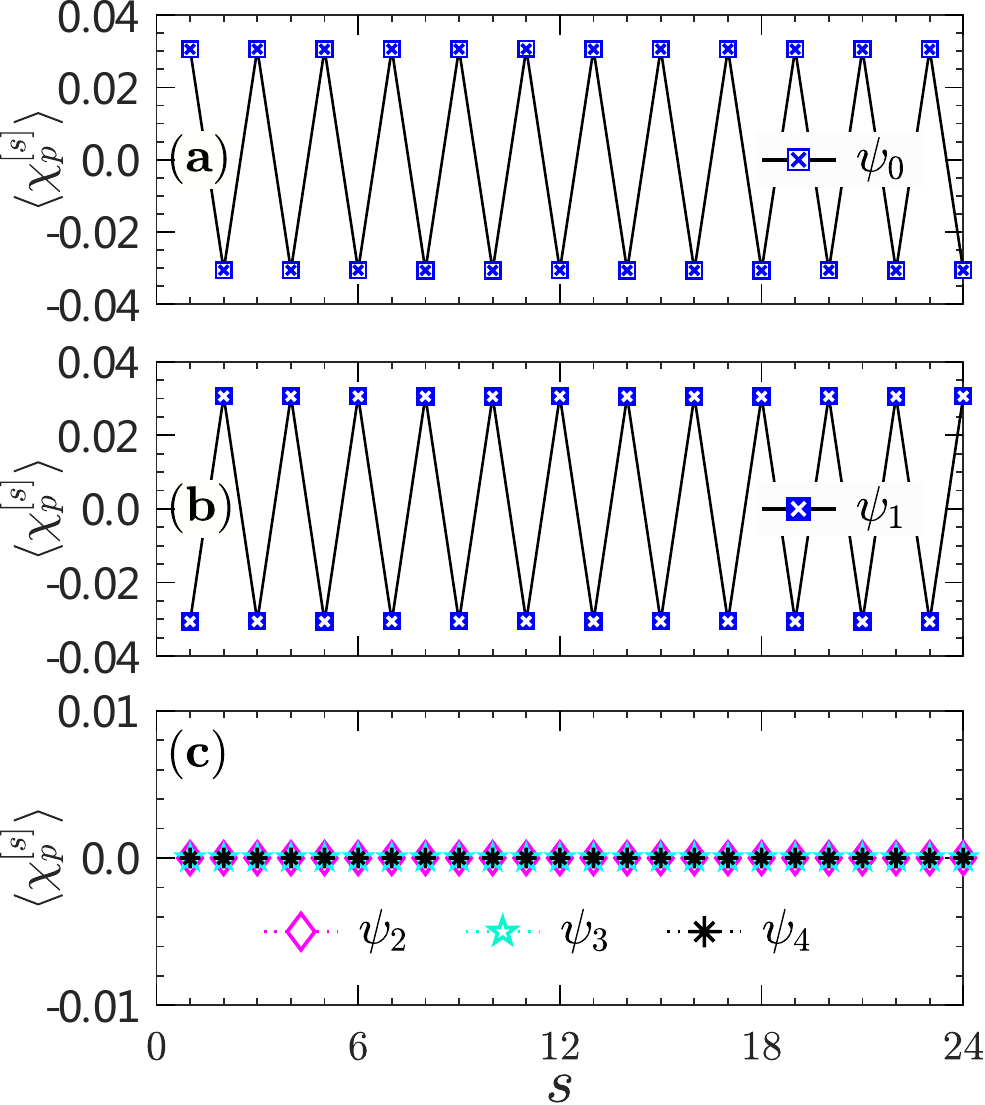}\\
\caption{Chirality distribution of the $\Gamma$-$\Gamma'$ model under a 24-site hexagonal cluster with $\Gamma'/\Gamma = 0.15$.
    (a) and (b) are for the two degenerate ground states while (c) is for higher excited states.}
    \label{FIGSM-Gp015Chi5}
\end{figure}

\vspace{-0.20cm}
\subsection{Hexagonal clusters with $\Gamma'/\Gamma$ = 0.15}\label{SMSec:Hexs}
In this section we focus on the \textit{quantum} $\Gamma$-$\Gamma'$ model with $\Gamma'/\Gamma = 0.15$
and discuss the properties of the emergent CS phase.
As demonstrated in the main text, the CS phase is known to have a doubly degenerate ground state.
Here we show that the two degenerate ground states have finite and opposing scalar chirality, while it is zero for higher excited states.
Figure~\ref{FIGSM-Gp015Chi5}(a) and (b) present the chiral distribution for the two (purified) degenerate ground states, respectively.
For each of the two, the chirality is staggered, consistent with Eq.~\eqref{SMEQ:ChiAmB}.
Moreover, their patterns are complementary and are reminiscent of the time reversal symmetry breaking.
Furthermore, for higher excited states (i.e., $\psi_2$, $\psi_3$, and $\psi_4$, etc.),
the scalar chirality is zero (see Fig.~\ref{FIGSM-Gp015Chi5}(c)).

\vspace{-0.20cm}
\subsection{Three- and four-leg tori/cylinders}
The twofold ground-state degeneracy is usually typical of hexagonal clusters but not the conventional tori,
which are at odds with the $C_{3v}$ symmetry of the model.
However, this symmetry should be plethoric and
it is not necessary to include all of its subgroups to ensure the degeneracy.
We find that the diamond-like tori of $2\times 3\times 3$ and $2\times 4\times 4$
could still hold a double degeneracy and
have the same energy spectra (at least for the first five energy levels)
to that of hexagonal clusters of $N$ = 18 and 32, respectively.
In general, the diamond-like tori of $2\times\!L\times\!L$ own the $C_2$ symmetry and mirror symmetry,
and these symmetries seem to be essential to maintain a twofold degenerate ground state.

\begin{figure}[!ht]
\centering
\includegraphics[width=0.90\columnwidth, clip]{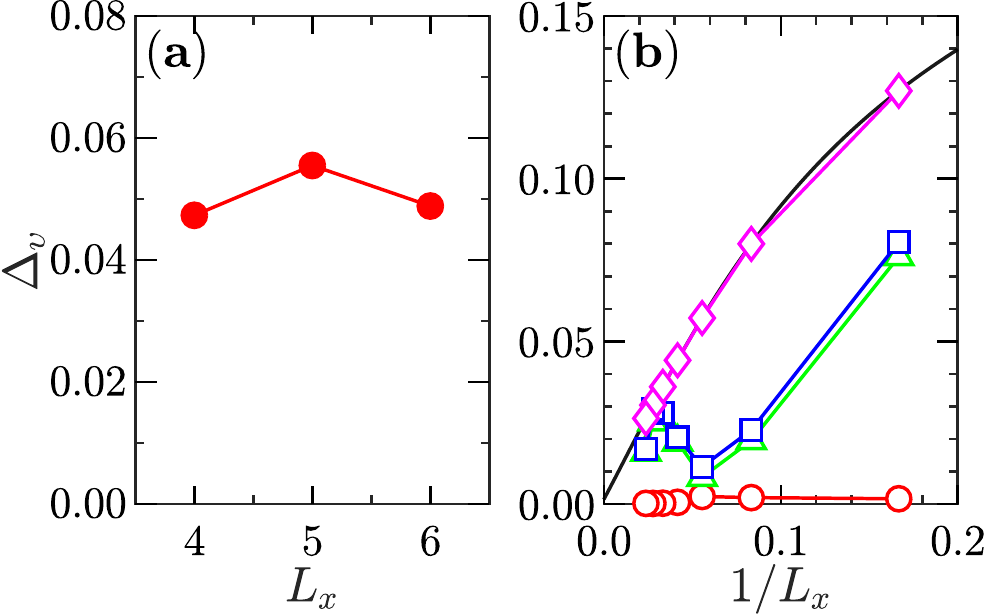}\\
\caption{Energy gaps of the $\Gamma$-$\Gamma'$ model under the $2\times\!L_x\times\!3$ clusters with $\Gamma'/\Gamma = 0.15$.
    (a) The lowest energy gap under the tori with $L_x$ = 4, 5, and 6.
    (b) The first four energy gaps under the cylinders with $L_x$ = 6, 12, 18, 24, 30, 36, and 42.
    The black solid line is the fitting of the fourth energy gaps $E_4$~(pink diamond).}
    \label{FIGSM-Gp015Gap}
\end{figure}

Nevertheless, the ground state is unique for general torus when $L_x \neq L_y$.
To show it, we calculate the energy gap on a sequence of $2\times\!L_x\times\!3$ tori
of length $L_x$ = 4, 5, and 6 (i.e., up to a 36-site cluster), see Fig.~\ref{FIGSM-Gp015Gap}(a).
the ground states are unique and are separated by a stable energy gap of roughly $\Delta \approx 0.05$.
It is arguable that the twofold degeneracy is likely to be achieved
by a $2\pi$-flux insertion which can pump an excitation from one edge to the other side.
Unfortunately, this procedure is not applicable to our model for the lack of $U(1)$ symmetry.

Because of the small clusters in Fig.~\ref{FIGSM-Gp015Gap}(a),
we can not conclude if it is gapless or not for infinite-size system.
To conquer this issues, we calculate the first few energy levels under cylinders of $2\times\!L_x\times\!3$,
where open boundary condition is applied along the $L_x$-direction.
Here, $L_x$ is chosen to be the multiply of six, i.e., $L_x$ = 6, 12, 18, 24, 30, 36, and 42.
As shown in Fig.~\ref{FIGSM-Gp015Gap}(b), the first energy gap $\Delta_{1}$ is very tiny,
while the second and third $\Delta_{2,3}$ are modest and are close in value.
The forth energy gap $\Delta_{4}$ is a little bit large and decreases monotonously with $L_x$.
It is interesting to note that $\Delta_{2,3}$ are not monotonically decreasing with $L_x$ but turn upward at $L_x = 18$.
They begin to go down again at $L_x = 36$ where their values are only slightly smaller than $\Delta_{4}$.
We also fit the fourth energy gaps $E_4$ and find that it vanishes when $L_x \to \infty$.
Given the abnormal behaviors of $\Delta_{2,3}$,
we speculate that there is a strong finite-size effect when $L_x \lesssim 18$.
This is partially verified by the entanglement entropy scaling shown in Fig.~\textcolor{red}{8}(b) in the main text,
from which we find that entanglement entropy $\mathcal{S}_{\rm vN}$ of clusters with $L_x > 18$ fall in a scaling formula.

\vspace{-0.30cm}
\section{Momentum-resolved exact-diagonalization calculation}\label{SMSEC:MomED}

Before discussing the momentum-resolved quantities we first enumerate the relevant symmetries and how they constrain the spin chirality.
With no external field we have
(i) time-reversal symmetry $\hat{T}$,
(ii) inversion symmetry $\hat{P}$ about the plaquette center,
(iii) $\hat{C}_{3c^*}$ about the normal $\hat{\bm{c}}^*$ to the honeycomb plane,
(iv) a $\hat{C}_{2b}$ about the crystallographic $\hat{\bm{b}}$ direction, and
(v) their combinations, including the mirror $\hat{M}_b = \hat{P}\hat{C}_{2b}$.
Among these, the symmetries involving the inversion operator exchange the $A$ and $B$ sublattices.
Note that none of the symmetries can interchange the $n$- and $p$-type chiral operators, so generically their expectation values will differ.
The sublattice transformation properties of the $\hat{\chi}_{ijk}^{\Delta}$ operators are summarized in Tab.~\ref{tab:sym}.

\begin{table}[h!]
\begin{ruledtabular}
  \begin{tabular}{c || c | c | c | c | c}
    $\hat{\bm{\mathcal{U}}}$ & $\hat{T}$ & $\hat{P}$ & $\hat{\Theta} = \hat{T}\hat{P}$ & $\hat{C}_{2b}$ & $\hat{M}_b = \hat{P}\hat{C}_{2b}$ \\
  \colrule
    $\hat{\bm{\mathcal{U}}}\hat{\bm{\chi}}^{\bm{A/B}}\hat{\bm{\mathcal{U}}}^{-1}$ & $-\hat{\chi}^{A/B}$ & $+\hat{\chi}^{B/A}$ & $-\hat{\chi}^{B/A}$ & $-\hat{\chi}^{A/B}$ & $-\hat{\chi}^{B/A}$ \\
  \end{tabular}
\end{ruledtabular}
\caption{Transformation properties of the chiral operators $\hat{\chi}_{ijk}^{\Delta}$ under select global and spatial symmetries.
  The spatial and $n/p$ indices have been suppressed to emphasize the sublattice structure.}
  \label{tab:sym}
\end{table}

\begin{table*}[htb]
\begin{ruledtabular}
  \begin{tabular}{c | c | c | c | c | c | c | c}
    $(\Gamma,\Gamma^{\prime}) = (1.00,0.15)$ & $\braket{\bm{S}_A}$ & $\braket{\bm{S}_B}$ & $\braket{\hat{W}_p}$ & $\braket{\hat{\chi}_n^A}$  & $\braket{\hat{\chi}_n^B}$ & $\braket{\hat{\chi}_p^A}$ & $\braket{\hat{\chi}_p^B}$ \\
    \colrule
    $\ket{\bm{\mathrm{K}_1}}$ & $-\num{0.219538}\ \hat{\bm{c}}^*$ & $+\num{0.219538}\ \hat{\bm{c}}^*$ & $-\num{0.230595}$ & $-\num{0.021406}$ & $+\num{0.021406}$ & $-\num{0.030583}$ & $+\num{0.030583}$ \\
    $\ket{\bm{\mathrm{K}_2}}$ & $+\num{0.219538}\ \hat{\bm{c}}^*$ & $-\num{0.219538}\ \hat{\bm{c}}^*$ & $-\num{0.230595}$ & $+\num{0.021406}$ & $-\num{0.021406}$ & $+\num{0.030583}$ & $-\num{0.030583}$ \\
    $\ket{\bm{\Gamma}}$ & $\phantom{+}\num{0.000000}\ \hat{\bm{c}}^*$ & $\phantom{+}\num{0.000000}\ \hat{\bm{c}}^*$ & $-\num{0.260523}$ & $\phantom{+}\num{0.000000}$ & $\phantom{+}\num{0.000000}$ & $\phantom{+}\num{0.000000}$ & $\phantom{+}\num{0.000000}$ \\
  \end{tabular}
\end{ruledtabular}
\caption{Momentum-resolved observables in the CS phase on the 24-$C_3$ cluster in order of ascending energy.
  The spin and chirality operators include factors of $S = 1/2$, while the plaquette operator is scaled by $S^{-6} = 2^6$.
  Despite the finite momentum of the ground state, each quantity is found to be uniform across the cluster.
  We see that the expectation values of the chirality operators are in line with the relationships
  in Eq.~(\textcolor{red}{18}) and Eq.~(\textcolor{red}{19}) in the main text implied by Tab.~\ref{tab:sym}.}
  \label{tab:24C3obs}
\end{table*}

To determine the symmetries protecting the degeneracy between $\ket{\bm{\mathrm{K}_{1,2}}}$ we employ a number of translationally-symmetric perturbations.
A uniform field in generic directions fails to break the degeneracy between the levels, indicating that the ``sublattice symmetry''
between $A$ and $B$ is relevant.
In a fermionic system one may use a staggered chemical potential to break this, however with spins we must also break the time-reversal symmetry
leading to a staggered field $\hat{\bm{h}}_A = -\hat{\bm{h}}_B = \hat{\bm{h}}$.
This perturbation breaks the degeneracy for almost all directions except $\hat{\bm{h}} = \hat{\bm{b}}$ along the $\hat{C}_{2b}$ axis.
Finally, we consider a staggered field where $\hat{\bm{h}}_B$ is related to $\hat{\bm{h}}_A = \hat{\bm{h}}$ through $\hat{\Theta}\hat{C}_{2b} = \hat{T}\hat{P}\hat{C}_{2b}$, which breaks the rest of the point group symmetries.
Under such a field the ground state remains degenerate for generic directions of $\hat{\bm{h}}$,
from which we conclude that the combination of $\hat{\Theta}\hat{C}_{2b}$ and the translations protect it.

\begin{figure}[!ht]
\centering
\includegraphics[width=0.90\columnwidth, clip]{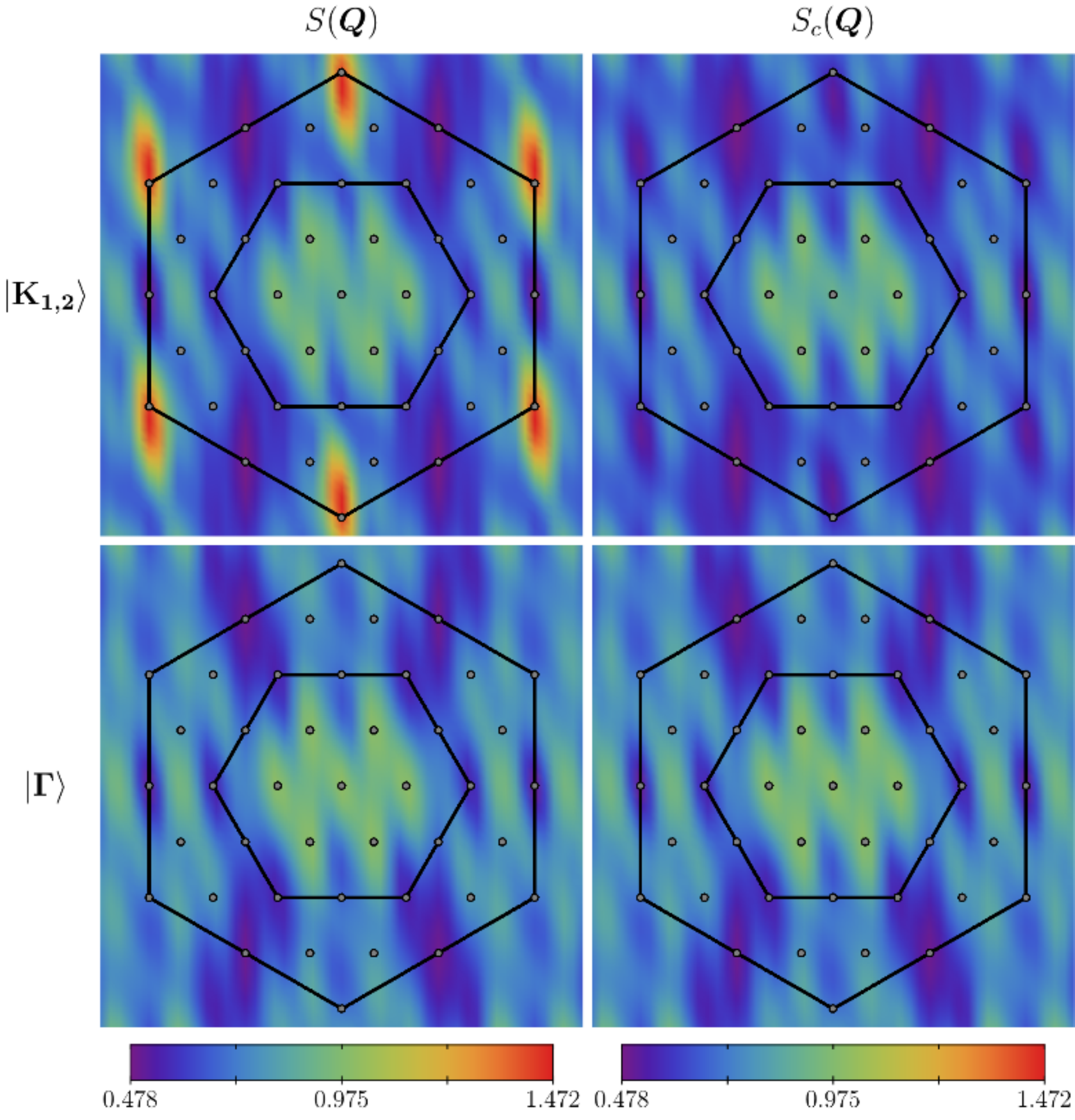}\\
\caption{Interpolated static spin structure factor on the 24-$C_3$ cluster in the CS phase at $(\Gamma,\Gamma^{\prime}) = (1.00,0.15)$ for the lowest lying three states.
  Dots within the first two Brillouin zones represent the accessible momenta.
  The first column uses the bare correlations $\braket{S_i^{\alpha}S_j^{\beta}}$ while the second subtracts off the static AFM moments via
  $\braket{S_i^{\alpha}S_j^{\beta}} - \braket{S_i^{\alpha}}\braket{S_j^{\beta}}$.
  Each plot is on the same colour scale.
  }\label{FIGSM-24C3SSQ}
\end{figure}

We now consider properties of the lowest excited states in the CS phase, including the doublet with momentum $\bm{q} = \bm{\mathrm{K}_{1,2}}$ on the 24-$C_3$ cluster.
There are two ways to obtain quantities resolved in each momentum sector.
One way is to use a small staggered pinning field at two sites to split the degeneracy, and take expectation values with those states.
Another is to find the matrix representations of the cluster translation operators $\hat{t}_{1,2}$ within the degenerate subspace $\braket{\Psi_n|\hat{t}_{1,2}|\Psi_m}$,
which we simultaneously diagonalize to find the symmetric linear combinations.
The expectation value of the spin, plaquette, and chiral operators are summarized in Tab.~\ref{tab:24C3obs}.
As mentioned in the main text, we find an AFM pattern of the moments along the $\hat{\bm{c}}^*$ direction in the doublet, as well as uniform values of the chiral
expectations in line with the symmetry relations in Eq.~(\textcolor{red}{18}) and Eq.~(\textcolor{red}{19}) in the main text.
The bare spin correlations $\braket{S_i^{\alpha}S_j^{\beta}}$ in the ground state yields a static structure factor with  a peak at $\bm{Q} = \bm{\Gamma^{\prime}}$ as shown in the top left corner of Fig.~\ref{FIGSM-24C3SSQ}.
After subtracting the uniform AFM moment the residual correlations lead to a static structure factor shown in the top right corner Fig.~\ref{FIGSM-24C3SSQ}, resembling that of the nearby $\ket{\bm{\Gamma}}$ state.
Since the DMRG magnetization scales to zero for large system sizes, we expect $S_c(\bm{Q})$ to be more representative of the correlations.

Finally, we look for evidence of topological order in the CS phase by analyzing the entanglement of the low-lying states.
ED calculations observe three clustered states which we assume to make up a quasi-degenerate subspace.
This assumption combined with the result of $c \approx 1$ from DMRG calculations suggests that semion topologial order with a two-fold topological degeneracy is possible.
We have considered the combinations $\{\ket{\bm{\mathrm{K}_1}},\ket{\bm{\mathrm{K}_2}}\}$, $\{\ket{\bm{\Gamma}},\ket{\bm{\mathrm{K}_1}}\}$, and $\{\ket{\bm{\Gamma}},\ket{\bm{\mathrm{K}_2}}\}$ as possible members of the quasi-degenerate subspace on the 24-$C_3$ cluster.
By minimizing the entanglement entropy of their linear combinations on each cycle of the torus we obtain the minimally entangled states.
The unitary transformation from the minimally entangled states on one cycle of the torus to another represents the modular $\mathcal{S}$-matrix.
In each case we find that the modular matrices are nearly identical for the different entanglement cuts, with the momentum states having minimal entropy.
This yields $\mathcal{S} \approx \mathbbm{1}$, indicating no topological order based on the 24-site calculation.


%



\end{document}